\documentclass[traditabstract]{aa} 
\usepackage{graphicx}
\usepackage{subfig}
\usepackage{txfonts}
\usepackage{natbib}
\usepackage{url}
\bibpunct{(}{)}{;}{a}{}{,} 
\usepackage{enumitem}

\usepackage{color}

\begin{document}

   \title{Constraining the escape fraction of ionizing photons from \ion{H}{ii} regions within NGC 300: A concept paper
    \thanks{Based on observations made with the NASA/ESA Hubble Space Telescope, and obtained from the Hubble Legacy Archive, which is a collaboration between the Space Telescope Science Institute (STScI/NASA), the Space Telescope European Coordinating Facility (ST-ECF/ESA) and the Canadian Astronomy Data Centre (CADC/NRC/CSA).}
\thanks{Based on observations made with ESO telescopes at the La Silla Paranal Observatory under program ID 077.B-0269(A).       
    }}

   \author{F. Niederhofer 
                        \inst{1,2,3}
                        \and
                        M. Hilker
                        \inst{3,4}
                        \and
                        N. Bastian
                        \inst{5}
                        \and
                        B. Ercolano
                        \inst{2,3} 
            }

   \institute{Space Telescope Science Institute, 3700 San Martin Drive, Baltimore, MD 21218, USA \\
   \email{fniederhofer@stsci.edu}
   \and
           Universit\"ats-Sternwarte M\"unchen, Scheinerstra\ss e 1, D-81679 M\"unchen, Germany
    \and
    Excellence Cluster Origin and Structure of the Universe, Boltzmannstr. 2, D-85748 Garching bei M\"unchen, Germany     
    \and       
   European Southern Observatory, Karl-Schwarzschild-Stra\ss e 2, D-85748 Garching bei M\"unchen, Germany
          \and
           Astrophysics Research Institute, Liverpool John Moores University, 146 Brownlow Hill, Liverpool L3 5RF, UK
              }


  \abstract
{
Using broadband photometry from the \textit{Hubble Space Telescope} in combination with \textit{Very Large Telescope} narrowband H$\alpha$ observations of the nearby spiral galaxy NGC~300, we explore a method for estimating the escape fractions of hydrogen-ionizing photons from \ion{H}{ii} regions within this galaxy. Our goal in this concept study is to evaluate the spectral types of the most massive stars using the broadband data and estimating their ionizing photon output with the help of stellar atmosphere models. A comparison with the H$\alpha$ flux that gives the amount of ionized gas in the \ion{H}{ii} region provides a measure of the escape fraction of ionizing photons from that region. We performed some tests with a number of synthetic young clusters with varying parameters to assess the reliability of the method. However, we found that the derived stellar spectral types and consequently the expected ionizing photon luminosity of a region is highly uncertain. The tests also show that on one hand we tended to overestimate the integrated photon output of a region for young ages and low numbers of stars, and on the other hand we mostly underestimated the combined ionizing luminosity for a large stellar number and older cluster ages.
We conclude that the proposed method of using stellar broadband photometry to infer the leakage of ionizing photons from \ion{H}{ii} regions is highly uncertain and dominated by the errors of the resulting stellar spectral types. Therefore this method is not suitable. Stellar spectra are needed to reliably determine the stellar types and escape fractions. Studies to this end have been carried out for the Magellanic Clouds.

}

\titlerunning{Escaping NGC 300}
   \maketitle
%
\section{Introduction}

Massive stars are the main source of radiation with energies high enough to ionize neutral H atoms. Most of this Lyman-continuum (LyC) emission is absorbed very quickly by neutral gas, thereby heating the surrounding interstellar medium (ISM).  However, some parts of the created \ion{H}{ii} region might be optically thin, and a certain fraction of the ionizing radiation eventually escapes into the diffuse medium of the galaxy or into the intergalactic medium (IGM). The escape fraction of ionizing photons, $f_{esc}$, affects the energy budget of the ISM as well as the evolution of a galaxy. 

It is not possible to directly determine $f_{esc}$ because any LyC radiation that is able to escape its parent system will be absorbed by the ISM of the Milky Way disk. For this reason, several observational studies used various methods and techniques to estimate the escape fraction of \ion{H}{ii} regions or entire galaxies. 
\citet{Leitherer95} measured the redshifted LyC flux of a sample of starburst galaxies (z$>$0.015) that is not absorbed by our Galaxy to infer $f_{esc}$. They found that the leakage of ionizing photons from these galaxies is below 3\%. Similar results for moderate redshifts (z$<$1.5) are found, for example, by \citet{Siana10}, \citet{Leitet13} and, very recently, \citet{Rutkowsky16}. 

\citet{Pellegrini12} used the method of ionization-parameter mapping to determine the optical depth of \ion{H}{ii} regions in the Large and Small Magellanic Cloud (LMC and SMC). Radiation-bound regions are characterized by transition regions in their ionization structure and can be measured by emission line ratios. \citet{Pellegrini12} used the [\ion{S}{ii}] to [\ion{O}{iii}] ratio for their analysis. They found that the most luminous \ion{H}{ii} regions are optically thin to LyC radiation and that regions with low gas column density and irregular shape tend to have higher escape fractions. 
The global fraction of ionizing photons that are leaving the Magellanic Clouds are determined to be between 10 and 20\%. 
\citet{Zastrow13} mapped the ionization structure of nearby dwarf starburst galaxies to analyze their amount of escaping ionizing radiation. Some galaxies in their sample show evidence for LyC photons leaking in a preferred limited direction.

Another method for inferring the escape fraction of \ion{H}{ii} regions in nearby galaxies includes the analysis of the high-mass stars within the regions.
With the knowledge of the spectral types of the stars responsible for the ionizing radiation, it is possible to estimate the expected total ionizing photon output $Q^0$ of these stars using theoretical atmosphere grids. A comparison with the total amount of the ionized gas (N$_{\mathrm{\ion{H}{ii}}}$) in the surrounding regions of these stars, using H$\alpha$ imaging, for example, then allows for an estimate of $f_{esc}$. \citet{Oey97} applied this method to \ion{H}{ii} regions within the LMC using available spectral classifications of O-type stars. Their analysis shows that a considerable fraction of the studied regions is optically thin to ionizing radiation with escape fractions of up to 50\%. This work has been revised by \citet{Voges08}. Using newer stellar atmosphere models (e.g., from \citealt{Smith02}), they found, in contrast to \citet{Oey97}, that the majority of the \ion{H}{ii} regions in the LMC is radiation bounded. In a similar work, \citet{Doran13} estimated the escape fraction of 30 Doradus to be between 0 and 60\%.

In this paper, we examine a similar approach to constrain the escape fraction from individual \ion{H}{ii} regions within the nearly face-on spiral galaxy NGC~300. Our method is based on inferring the parameters of the hottest stars within the emission nebulae with the help of broadband photometric data to determine whether photometric classification is accurate enough. The implication
is that if this is the case, the method can be applied to many nearby galaxies without the need of time-consuming spectroscopic observations. It is expected that the escape fraction from H\textsc{ii} regions varies within galaxies. As the gas densities increase toward the centers, regions in the inner parts of galaxies may have lower escape fractions locally than regions in the outer parts. Similarly, in the outskirts the chances are highest that ionizing photons can escape the galaxy, so that young clusters away from the central regions may preferentially contribute to the photons ionizing the intergalactic medium. \citet{ConroyKratter12} adopted a similar idea and proposed massive runaway stars which escape the high-density regions and enhance $f_{esc}$ from galaxies in the early Universe.

For the method described in this paper, we used archival photometric data from the \textit{Hubble Space Telescope (HST)} in the filters $F475W$, $F606W,$ and $F814W$, taken with the Advanced Camera for Surveys \citep{Dalcanton09,Gogarten10}. Additionally, we used  narrowband H$\alpha$ images observed at the \textit{Very Large Telescope (VLT)} with the FORS2 instrument \citep{Bresolin09} to estimate the amount of ionized gas.

The structure of the paper is the following: in Sect. \ref{sec:obs} we describe the observations and the data reduction procedure. The method used to derive the escape fractions and the results are  explained in Sect. \ref{sec:res}. The discussion and final conclusions are presented in Sect. \ref{sec:disc}.

\section{Observations and data reduction\label{sec:obs}}

In this section we describe the data sets that we used to determine the escape fraction of a sample of \ion{H}{ii} regions within NGC~300. The first set comprises photometric broadband data in three filters taken with \textit{HST}. We used this photometry to obtain the temperatures and ionizing fluxes $Q^0$ of the stars in the \ion{H}{ii} regions.
The second data set consists of narrowband H$\alpha$ images taken with the \textit{VLT}/FORS2 instrument. The Balmer line flux is a measure of the amount of ionizing stellar photons that are absorbed by the surrounding material.

\subsection{Hubble dataset\label{sec:hst}}

The broadband photometric data of NGC~300 were taken with the Advanced Camera for Surveys (ACS) onboard the \textit{Hubble Space Telescope (HST)}. The galaxy was observed November 08 - 11~2006 as part of the ACS Nearby Galaxy Survey Treasury (ANGST) program \citep{Dalcanton09}. Three filters were used to observe NGC~300: $F475W$, $F606W,$ and $F815W$. We used three slightly overlapping fields that cover a radial strip of NGC~300 extending from the northwest to the center of the galaxy (see Figs. 1 and 2 in \citealt{Gogarten10}). Each of these fields covers an area of $\sim$3$\farcm$8 x 3$\farcm$8. They are referred to as Wide 1, Wide 2, and Wide 3 (going from the outskirts to the center). The pixel scale of the ACS is 0$\farcs$05 per pixel, corresponding to $\sim$0.47~pc at the distance of NGC~300. We assumed a distance of 1.93~Mpc \citep{Bresolin05}. For details of the data reduction see \citet{Dalcanton09}, and \citet{Gogarten10} for a description of the photometry and its use in constraining the star formation history (SFH) of NGC~300. We used the fully reduced photometric tables that are available for the public on the ANGST website\footnote{\url{http://archive.stsci.edu/prepds/angst/index.html}}.

\subsection{VLT/FORS2 dataset\label{sec:fors2}}
Our second dataset consists of archival narrowband H$\alpha$ images of NGC~300 (\citealt{Bresolin09}, under the program
077.B-0269(A)).
The data have been observed in 2006, June 22 and July 1 with the FORS2 instrument mounted at ESO's (European Southern Observatory) \textit{Very Large Telescope} on Cerro Paranal. The images consist of three different pointings and cover the northwestern, central, and eastern part of the galaxy. The seeing during the first night of observations was between $\sim$0$\farcs$7 and 0$\farcs$9, thus corresponding to $\sim$7.5 - 8.4~pc at the assumed distance to NGC~300. In the second night the seeing was worse with values between $\sim$1$\farcs$0 and 1$\farcs$6, corresponding to $\sim$9.4 - 15.0~pc. The H$\alpha$ filter used in combination with the SR collimator has a central wavelength of 656.3~nm.

The flat-field correction and the bias subtraction were made using the standard ESO pipeline EsoRex. For the further processing of the data we used IRAF\footnote{IRAF is distributed by the National Optical Astronomy Observatories, which is operated by the Association of Universities for Research in Astronomy, Inc., under cooperative agreement with the national Science Foundation.}. Cosmic rays were removed with the task \texttt{L.A.Cos} \citep{vanDokkum01}. After this, the images of the same fields were transformed to the same reference point using \texttt{geomap} and \texttt{gregister} and averaged together with the task \texttt{imcombine,} thereby blurring the images from the first night to the resolution of the images from the second night using the IRAF task \texttt{gauss}. As the H$\alpha$ image does not only contain flux from the H$\alpha$ line but also emission from the stellar continuum, this stellar component had to be removed from the science data. The stellar continuum was subtracted with the help of the narrowband filter H$\alpha$/4500, which is centered off the H$\alpha$ emission line at 666.5~nm. The data in this filter were taken in the second observation night and were processed in the same way as the actual science data. As a first step in the continuum subtraction, the H$\alpha$ image was blurred using the IRAF task \texttt{gauss} to fit the resolution of the image in the H$\alpha$/4500 filter because these observations were taken at systematically larger seeings. Next, all stars in both images were identified with the task \texttt{daofind} of the DAOPHOT package \citep{Stetson87}, and an aperture photometry was performed on them using \texttt{phot}. The tables of the stars were cross correlated to find stars that were detected in both images. After this, we plotted the counts of the stars in the H$\alpha$ image versus the stellar counts in the H$\alpha$/4500 image. The slope of the least-squares-fitted line to the data points yields the scaling factor between the images in the two filters. Finally, we subtracted the scaled H$\alpha$/4500 image from the science data to remove the stellar continuum.
       
For an absolute flux calibration the star \textit{Feige 110} served as the photometric standard \citep{Oke90}. The conversion from instrumental counts into physical units (erg~s$^{-1}$~cm$^{-2}$) was made as described in \citet{Jacoby87}. The conversion factor $S$ is given by 
\begin{equation}
S = \frac{t_{exp}\int F_{\lambda}T_{\lambda} \mathrm{d}\lambda}{C_{inst} 10^{0.4k_{\lambda}x}} 
\label{eq:conv}
,\end{equation}

where $t_{exp}$ is the exposure time, $F_{\lambda}$ the theoretical flux of the standard star, $T_{\lambda}$ the transmission curve of the filter, $C_{inst}$ the counts in the image, $k_{\lambda}$ the extinction coefficient, and $x$ the airmass.
To convert the counts into the actual H$\alpha$ flux that arrives at the telescope, the transmission curve of the used filter has to be taken into account and the exact shape of the emission line spectrum must be known. As a good approximation, we here
assumed that the H$\alpha$ line spectrum is a $\delta$ peak that is redshifted by 145~km~s$^{-1}$ (the heliocentric radial velocity of NGC~300) with respect to the rest frequency. Therefore, the conversion between the observed counts and the flux of the emission using the conversion factor $S$ from equation \ref{eq:conv} is given by
\begin{equation}
F_{\mathrm{H}\alpha}=\frac{S~C_{inst}~10^{0.4k_{\mathrm{H}\alpha}x}}{T_{\mathrm{H}\alpha}}
\label{eq:HaFlux}
,\end{equation}
where $T_{\mathrm{H}\alpha}$ is the filter transmission at the position of the H$\alpha$ line.

However, the narrowband H$\alpha$ filter contains not only the H$\alpha$ line, but also the nearby [\ion{N}{ii}] $\lambda$6548 and $\lambda$6583 lines. As the [\ion{N}{ii}]/H$\alpha$ ratio is not known for all the regions, it is not possible to correct for the additional transmitted [\ion{N}{ii}] lines for each region individually. Instead we assumed that the [\ion{N}{ii}] emission is 0.1 of the total H$\alpha$+[\ion{N}{ii}] emission, as was done by \citet{Roussel05} for NGC 300 \ion{H}{ii} regions, and in this way removed the blending 
flux. Typically, we expect a scatter of $\pm$0.04 for this ratio (cf. \citealt{Bresolin09}).

\section{Methods and results\label{sec:res}}

\begin{table} 
\caption{NGC 300 Parameters\label{tab:NGC300_Param}}
\centering
\begin{tabular}{l c } 
\hline\hline
\noalign{\smallskip}
Parameter & Value\\
\noalign{\smallskip}
\hline
\noalign{\smallskip}
R.A. (J2000.0) & 00:54:53.54\tablefootmark{a} \\
Dec (J2000.0) & $-$37:41:04.3\tablefootmark{a} \\
Morphological type & Scd\tablefootmark{a} \\
R$_{25}$ & 19$\farcm$50\tablefootmark{a} \\ 
Distance & 1.93 Mpc\tablefootmark{b} \\
Corresponding distance modulus & 26.43 mag \\
Galactic foreground extinction (A$_{\rm{V}}$) & 0.035 mag\tablefootmark{c} \\
Heliocentric radial velocity & 145 km~s$^{-1}$~\tablefootmark{a} \\
\noalign{\smallskip}
\hline

\tablefoottext{a}{HyperLeda database \citep{Paturel03}} \\
\tablefoottext{b}{\citet{Bresolin05}} \\
\tablefoottext{c}{\citet{Schlafly11}} \\
\end{tabular}
\end{table}

\subsection{Basic steps}
We used broadband \textit{HST} images together with narrowband H$\alpha$ data taken at the \textit{VLT} with the FORS2 instrument to examine the escape fraction $f_{esc}$ of \ion{H}{ii} regions across the nearby Sculptor group spiral galaxy NGC~300. The parameters of this galaxy 
are summarized in Table~\ref{tab:NGC300_Param}. The H$\alpha$ flux of \ion{H}{ii} regions is a measure of how much of the ionizing flux of the hot stars located inside the region is absorbed by the surrounding medium. Comparing this quantity with the actual integrated ionizing photon emission rate $Q^0$ of all stars inside the \ion{H}{ii} region yields the escape fraction of that region. Assuming Case B recombination and a typical electron temperature of 10,000~K, the LyC to H$\alpha$ photon emission rate conversion factor is 2.2 \citep{Hummer87}, which translates into a conversion of H$\alpha$ luminosity to LyC photons $Q(\mathrm{H\alpha})$ of $Q(\mathrm{H\alpha})=7.31\times 10^{11}L(\mathrm{H\alpha})s^{-1}$ (\citealt{Kennicutt95}, \citealt{Kennicutt98}). $Q(\mathrm{H\alpha})$ gives the number of LyC photons emitted by the stars that are absorbed and create the observed H$\alpha$ luminosity $L(\mathrm{H\alpha})$.
The escape fraction is therefore given by
\begin{equation}
f_{esc}=\frac{Q^0-Q_{\mathrm{H\alpha}}}{Q^0}
\label{eq:Fesc}
.\end{equation}

We determined the escape fraction of individual \ion{H}{ii} regions in three  basic steps:
\begin{enumerate}
\item Measure the H$\alpha$ flux of individual \ion{H}{ii} regions from the FORS2 data set as described in Sects. \ref{sec:fors2} and \ref{sec:Ha_flux}.
\item Estimate the integrated ionizing photon flux $Q^0$ of a region by converting the magnitudes and colors from stellar photometry described in Sect. \ref{sec:hst} into the stellar effective temperatures $T_{eff}$ and surface gravity log~G and the visual extinction $A_V.$ 
\item Calculate the escape fraction by using Eq. \ref{eq:Fesc}.
\end{enumerate}

More details about the methods employed are given below.

\subsection{Measure of the H$\alpha$ flux\label{sec:Ha_flux}}

To determine the integrated flux of the \ion{H}{ii} regions in the H$\alpha$ image, we made use of the \textit{CASAviewer} of the common astronomy software applications (CASA) package, which is originally designed to analyze astronomical radio data-cubes. For each region we first determined the local sky background and its standard deviation $\sigma$ in a polygonal area with no flux around the region. We then summed the counts inside the isophotal contour that is at 4$\sigma$ above the sky background and converted the integrated and background-subtracted counts into actual flux using Eq. \ref{eq:HaFlux} and considering the additional [N\textsc{ii}] component. Finally we corrected the flux for extinction, which is determined in the next step. To correct the H$\alpha$ flux, we took into account that the typical ratio of H$\alpha$ to stellar extinction at 656.3~nm equals 1.5 \citep{Calzetti01}.

\subsection{Determination of the stellar parameters}

Based on the three-filter broadband photometry, we determined $T_{eff}$ and log~G of the hot stars inside our \ion{H}{ii} sample plus the visual extinction $A_V$. This was done in two steps. We assumed that all the hot stars inside a single \ion{H}{ii} region have the same age and that all stars in one \ion{H}{ii} region have the same extinction and a Cardelli extinction law \citep{Cardelli89} with an R$_{\mathrm{v}}$ of 3.1. Using these constraints, we fit a single theoretical isochrone to the color-magnitude
diagram (CMD) of each stellar population in an \ion{H}{ii} region. 
Our method of fitting the isochrone was to minimize the geometrical distance in the CMD of an isochrone of given age and extinction to each star. To account for different numbers of stars in different parts of the CMD, we weighted each distance by the inverse number of stars in a 0.5 magnitude bin. 
For the fit we used all stars that form the main sequence (MS) down to a magnitude of $\sim$25 - 26. We then excluded from the fit stars that clearly deviated from the MS. 
The theoretical isochrones are from the Padova isochrone set \citep{Marigo08}. As a distance to NGC~300 we adopted 1.93~Mpc \citep{Bresolin05}, which translates into a distance modulus of 26.43~mag. For the metallicities of the stars we used the values give by \citet{Gogarten10}, which extend between [M/H]=$-$0.47 in the range 2.7~kpc $< r <$ 3.6~kpc and [M/H]=$-$0.24 in the central region for the youngest stars. In the second step we fit the stellar effective temperature and surface gravity using the \textup{\textit{\textup{tool for astrophysical data analysis}}} \textsc{ta-da} \citep{daRio12}. In \textsc{ta-da} we used the model grid from \citet{Marigo08} as the evolutionary model and the BT-Settl2010 grid \citep{Allard11} for the synthetic spectra.

\begin{figure}  
\centering
   \includegraphics[width = 8cm]{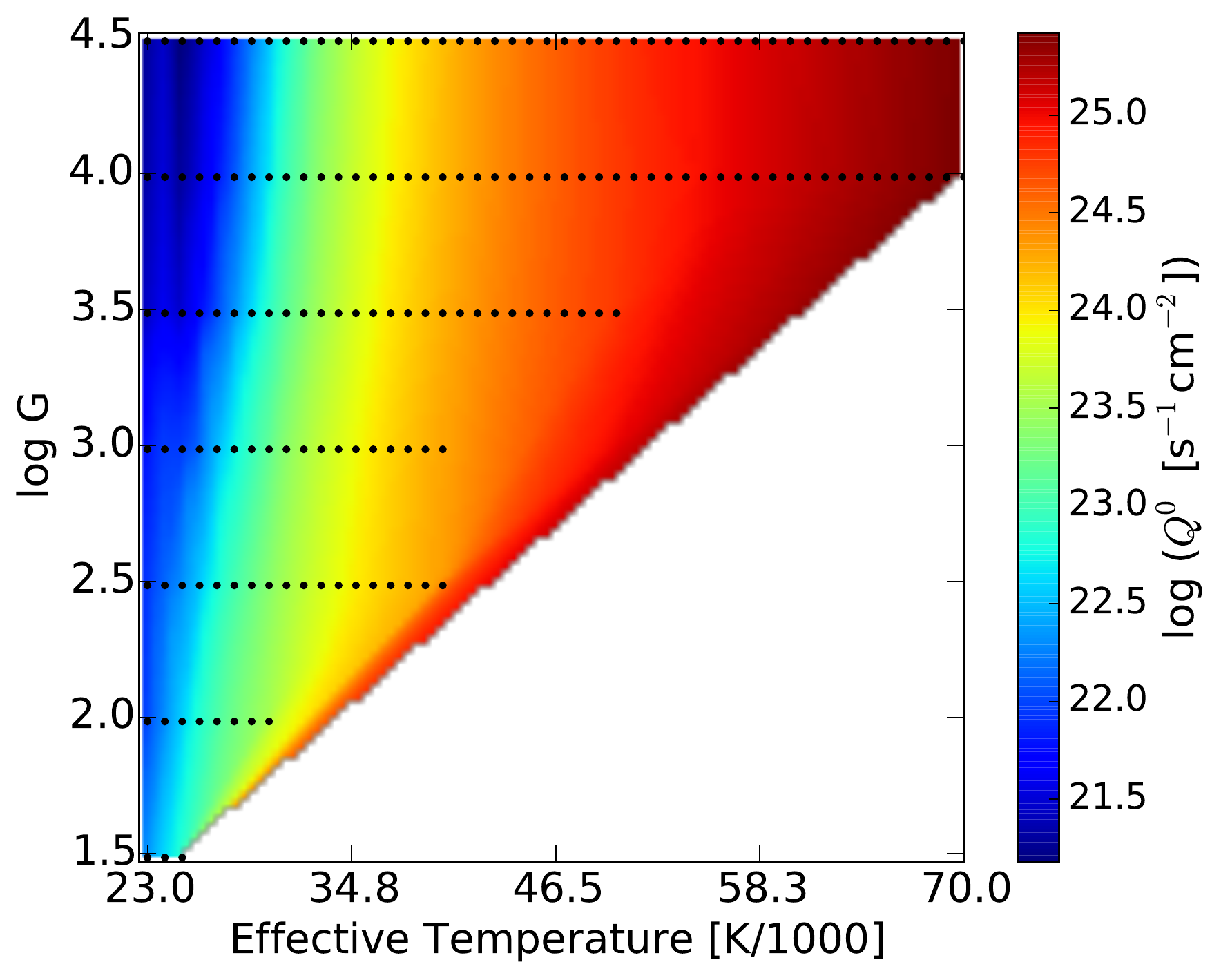}
   \caption{Ionizing photon flux as a function of the stellar effective temperature and surface gravity log~G from the BT-Settl2010 models for a metallicity of Z=0.006. The black dots show the positions of the individual calculated model points, while the colors represent the interpolated values. The color scale is in logarithmic units.}
              \label{fig:BT-Settl}
    \end{figure}

With the stellar parameters in hand, we calculated the ionizing photon output rate of the stars. For the sake of consistency, we used the BT-Settl2010 atmosphere model grid to calculate $Q^0$. For this we downloaded a grid of stellar spectra at various temperatures, surface gravities, and metallicities from the BT-Settl web page\footnote{\url{https://phoenix.ens-lyon.fr/Grids/BT-Settl/}}. To derive the ionizing photon flux of the model stars, we integrated the theoretical spectra up to the Lyman limit of 912~$\AA$ by
dividing each spectral bin by the photon energy at the respective wavelength. We then calculated the ionizing photon fluxes of the stars in the \ion{H}{ii} regions by interpolating the created grid of fluxes in temperature, surface gravity, and metallicity. Finally, we multiplied the resulting photon flux by the fitted surface of each star (also from \textsc{ta-da}) to obtain the total ionizing photon luminosity. As an example, Fig.~\ref{fig:BT-Settl} shows the rate of emitted LyC photons as a function of stellar temperature and surface gravity for a metallicity Z=0.006. The black dots indicate the positions of the calculated values from the BT-Settl2010 models, while the different colors are the interpolated values.

\begin{table} 
\caption{\ion{H}{ii} region sample\label{tab:HIIRegions}}
\centering
\begin{tabular}{l c c c c} 
\hline\hline
\noalign{\smallskip}
ID & R.A. & Dec & D$_{\mathrm{Center}}$ & ID\\
   & (J2000.0) & (J2000.0) & (kpc) & \citet{Deharveng88} \\
(1) & (2) & (3) & (4) & (5) \\   
\noalign{\smallskip}
\hline
\noalign{\smallskip}
1 &   0 54 24.87 & $-$37 39 47.1 & 3.26 & D15\\
2 &   0 54 25.63 & $-$37 39 09.8 & 3.27 & D17\\
3 &   0 54 27.26 & $-$37 40 01.6 & 2.97 & D20\\
4 &   0 54 28.77 & $-$37 38 36.6 & 3.07 & D25\\
5 &   0 54 29.45 & $-$37 37 23.3 & 3.38 & D26\\
6 &   0 54 31.66 & $-$37 37 51.0 & 3.02 & D30/D31\\
7 &   0 54 32.42 & $-$37 38 41.1 & 2.69 & D32/D34\\
8 &   0 54 35.25 & $-$37 39 42.6 & 2.16 & ...\\
9 &   0 54 35.33 & $-$37 39 35.7 & 2.18 & D37\\
10 &  0 54 36.06 & $-$37 39 50.6 & 2.05 & D38\\
11 &  0 54 41.84 & $-$37 38 36.4 & 1.89 & D48\\
12 &  0 54 42.19 & $-$37 39 02.7 & 1.69 & D49/D50\\
13 &  0 54 44.24 & $-$37 40 26.3 & 1.08 & D56\\
14 &  0 54 45.19 & $-$37 41 49.0 & 1.01 & ...\\
15 &  0 54 45.35 & $-$37 38 47.0 & 1.57 & D61\\
16 &  0 54 49.88 & $-$37 41 12.5 & 0.41 & D74\\
17 &  0 54 50.70 & $-$37 40 24.1 & 0.48 & D76\\
18 &  0 54 51.33 & $-$37 41 45.9 & 0.46 & D81\\
19 &  0 54 51.73 & $-$37 39 38.6 & 0.82 & D84\\
20 &  0 54 52.05 & $-$37 41 40.6 & 0.38 & D85\\

\noalign{\smallskip}
\hline

\end{tabular}
\tablefoot{ Column 1: identification of the \ion{H}{ii} regions, Col. 2: right ascension, Col. 3: declination, Col. 4: distance to the galactic center (assuming a distance to NGC~300 of 1.93~Mpc), Col. 5: identification from \citet{Deharveng88}.}
\end{table}

\begin{figure}
\centering
\begin{tabular}{l}
\includegraphics[width=8.6cm]{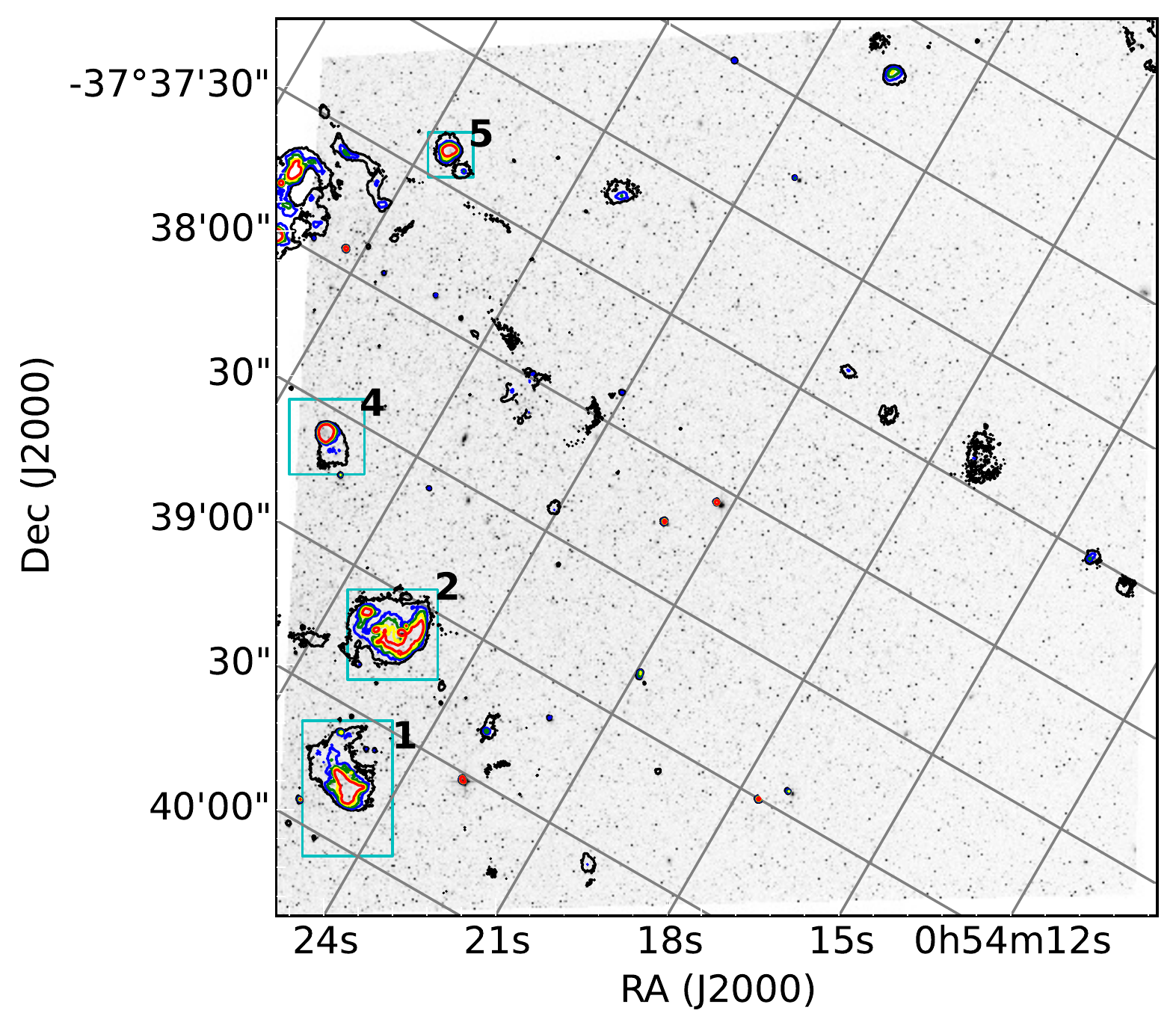} \\
\includegraphics[width=8.7cm]{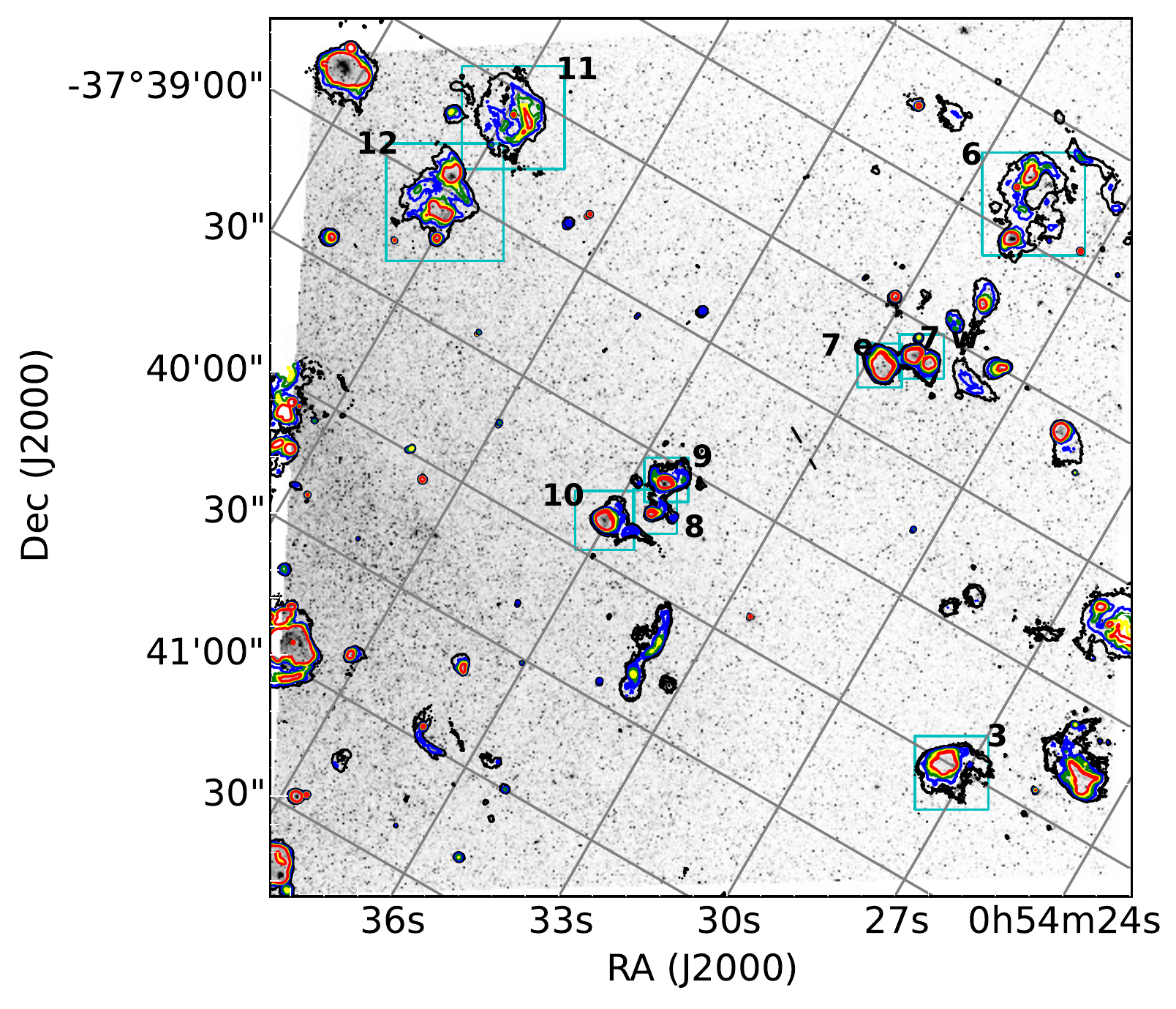} \\
\includegraphics[width=9.2cm]{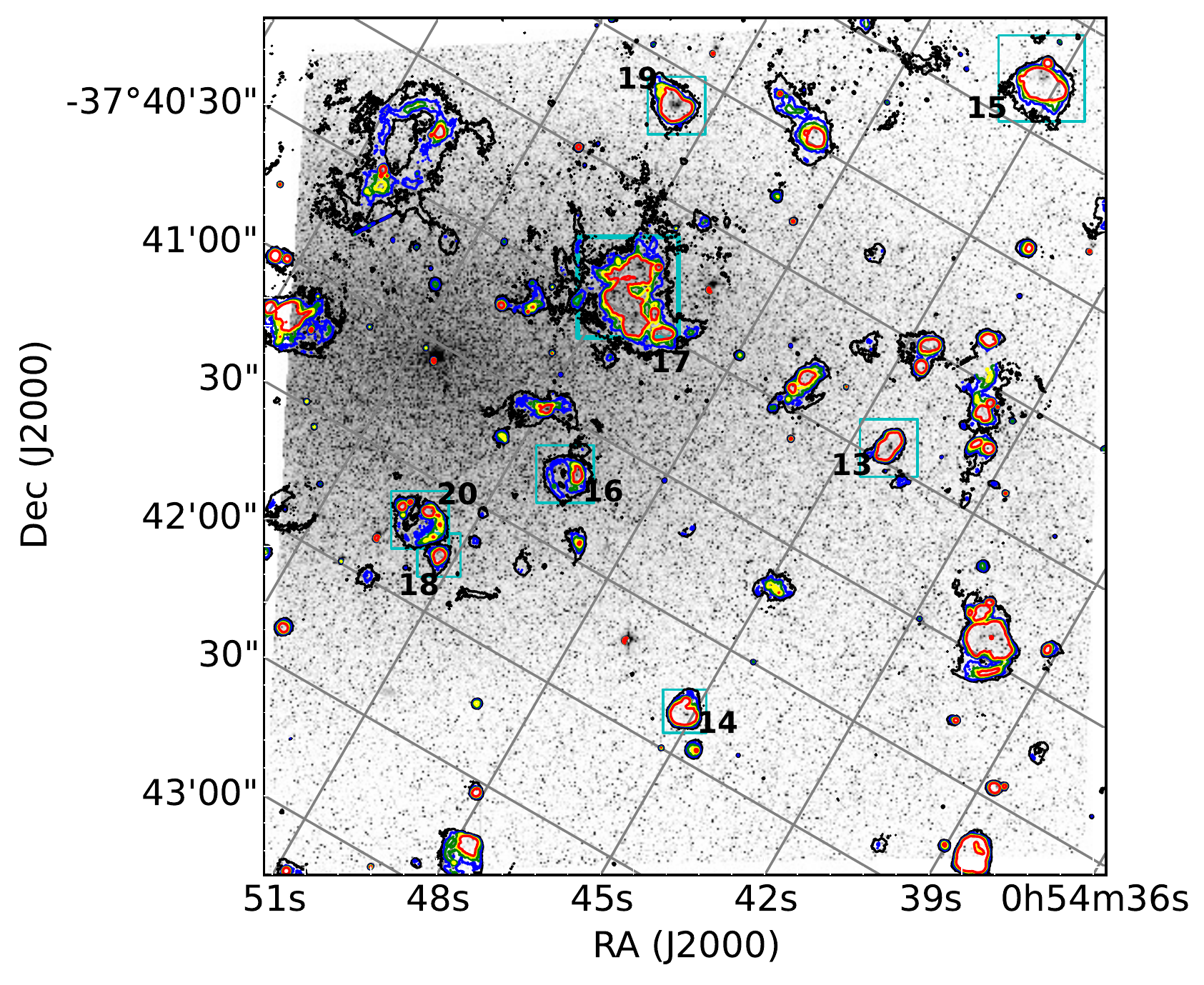} 
\end{tabular}
\caption{HST images of the three Wide fields with overplotted contours of H$\alpha$ emission (VLT/FORS2; \citealt{Bresolin09}). \textit{Top}: Wide 1, \textit{middle}: Wide 2, \textit{bottom}: Wide 3.}
\label{fig:HIIRegions}
\end{figure}

\subsection{Analysis and results}

We analyzed a sample of 20 \ion{H}{ii} regions of different sizes and shapes at various distances from the center of the galaxy (see Fig.~\ref{fig:HIIRegions}) with this method. Table~\ref{tab:HIIRegions} lists all the regions. We selected all objects that were classified as stars (object type $\leq$2) in the \citet{Dalcanton09} catalog that are located inside the compact regions of the \ion{H}{ii} regions because these stars are considered to be responsible for the ionization of their surrounding gas. We fit the parameters of all stars that lay inside pre-defined color ($F475W - F814W$ colors between $-$1 and 0.5) and magnitude (brighter than 25$\mathrm{^{th}}$ magnitude in the $F814W$ filter) cuts. This cuts restrict our fits to main-sequence stars hotter than $\sim$ 25,000~K and massive evolved stars. Given the logarithmic dependence of $Q^0$ on the temperature, the contribution of stars with temperatures $\leq$ 25,000~K is negligible. 
Unfortunately, many of the regions in our \ion{H}{ii} region sample are sparsely sampled in the upper parts of the CMD, and therefore fitting of an isochrone to it is ambiguous. For the further examination of our method, we selected a subsample of nine regions that are better sampled (see Fig.~\ref{fig:CMDs}). Table~\ref{tab:ExcellentRegions} lists the results of the calculated escape fractions of the shortlisted regions. 
On the one hand, we found for half of the regions "negative" escape fractions.
These values result from the fact that we measured more H$\alpha$ photons than can be explained by the integrated LyC luminosity of the stars within the \ion{H}{ii} region. This indicates that our fit underestimated the $Q^0$ flux, down to a factor of $\sim$10. On the other hand, we also obtained a quite high value for $f_{esc}$ of more than 60\%. These values are probably also very uncertain. As we discuss in the next section, our tests show that $Q^0$ can be overestimated by factors of more than four. 
In regions with negative values for $f_{esc}$ , much younger ages must be assumed to bring the integrated ionizing photon flux from the stars and the observed H$\alpha$ flux in agreement. In clusters where the upper parts of the CMDs are not populated, such as regions 4 or 13, younger ages would still be in agreement with the observations. In regions 2 or 3, however, the ages needed to explain the H$\alpha$ luminosity miss the bright evolved stars and therefore seem to disagree with the observations.

It is clear that our results are contaminated by uncertainties. To identify these errors and improve on the method, we performed additional tests, as described in the next section.

\begin{figure*}[htp]
  \centering
  
  \subfloat[\ion{H}{ii} region 2]{\label{figur:1_zoom}\includegraphics[width=64mm]{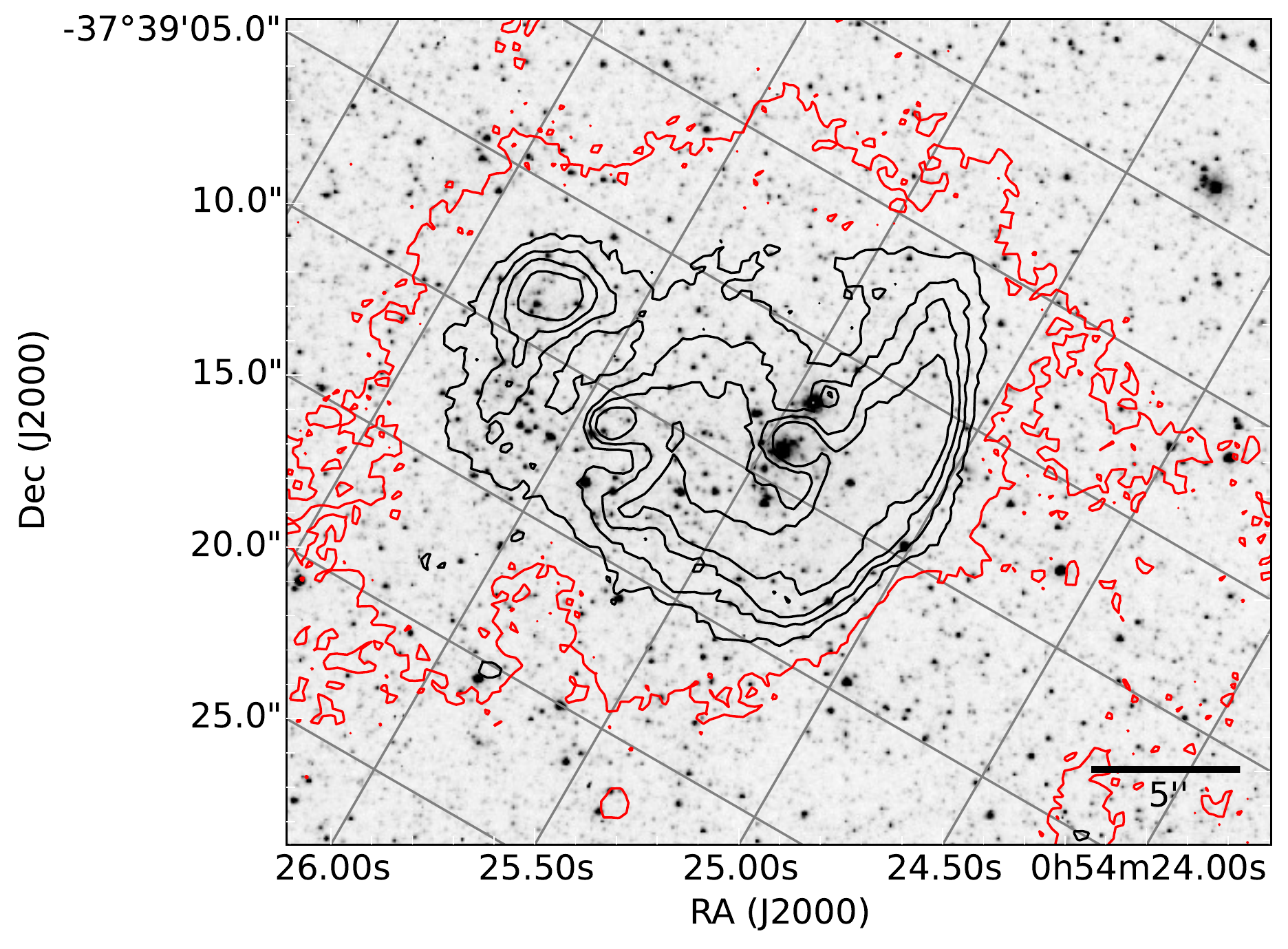}}
  \subfloat[\ion{H}{ii} Region 3]{\label{figur:2_zoom}\includegraphics[width=64mm]{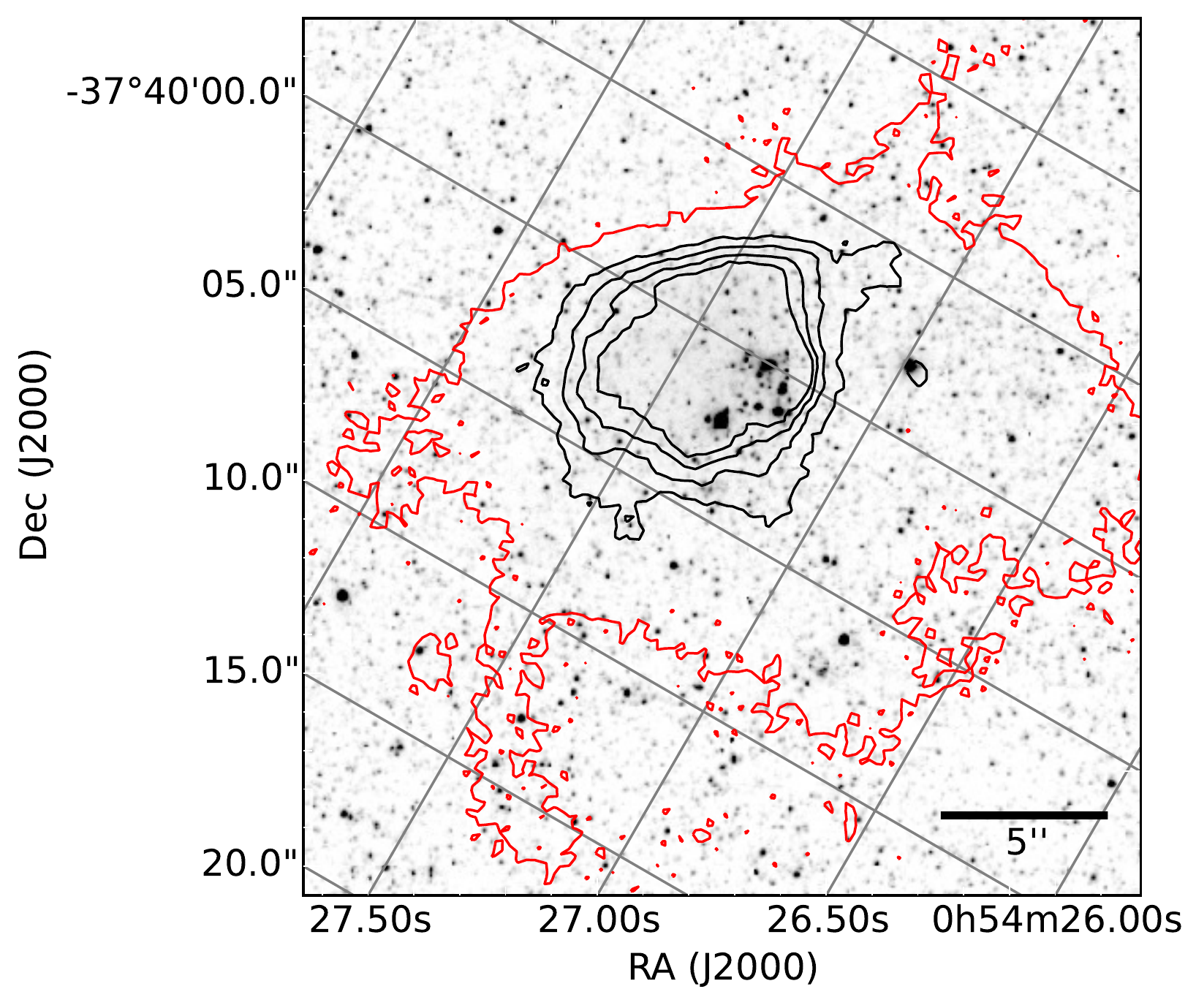}}
  \subfloat[\ion{H}{ii} Region 4]{\label{figur:3_zoom}\includegraphics[width=64mm]{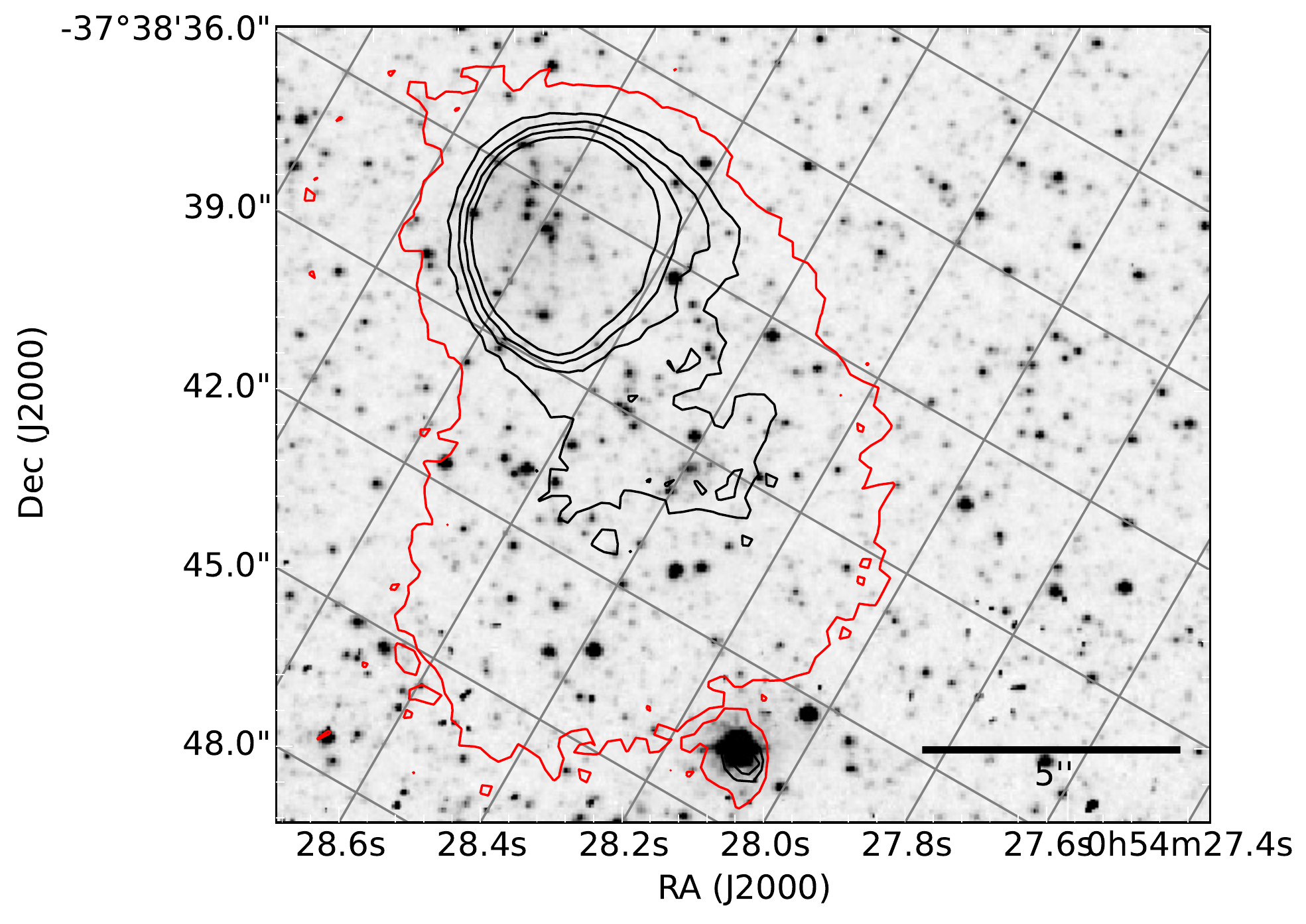}}
  \\
  \subfloat[\ion{H}{ii} region 6]{\label{figur:4_zoom}\includegraphics[width=64mm]{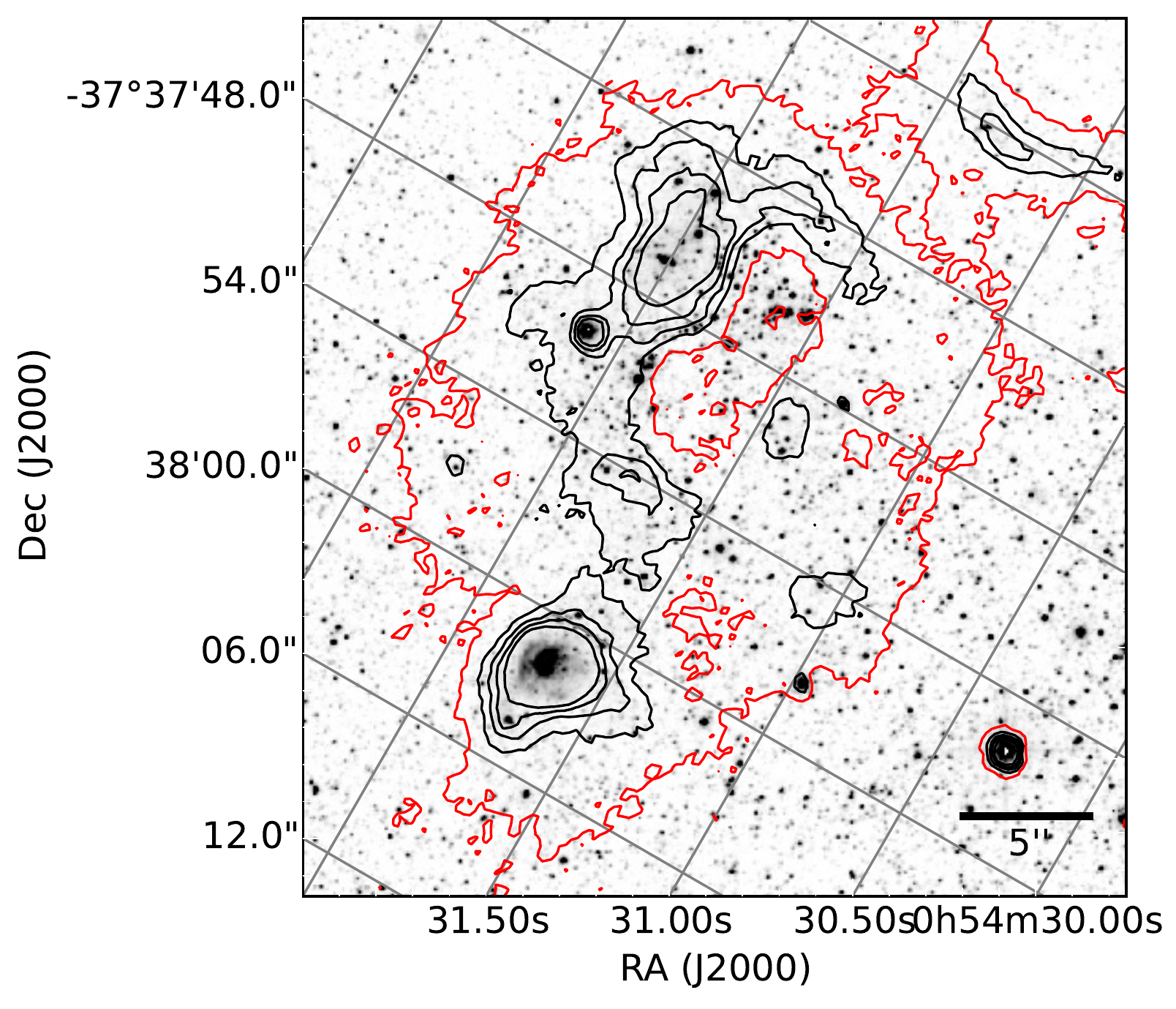}}
  \subfloat[\ion{H}{ii} Region 12]{\label{figur:5_zoom}\includegraphics[width=64mm]{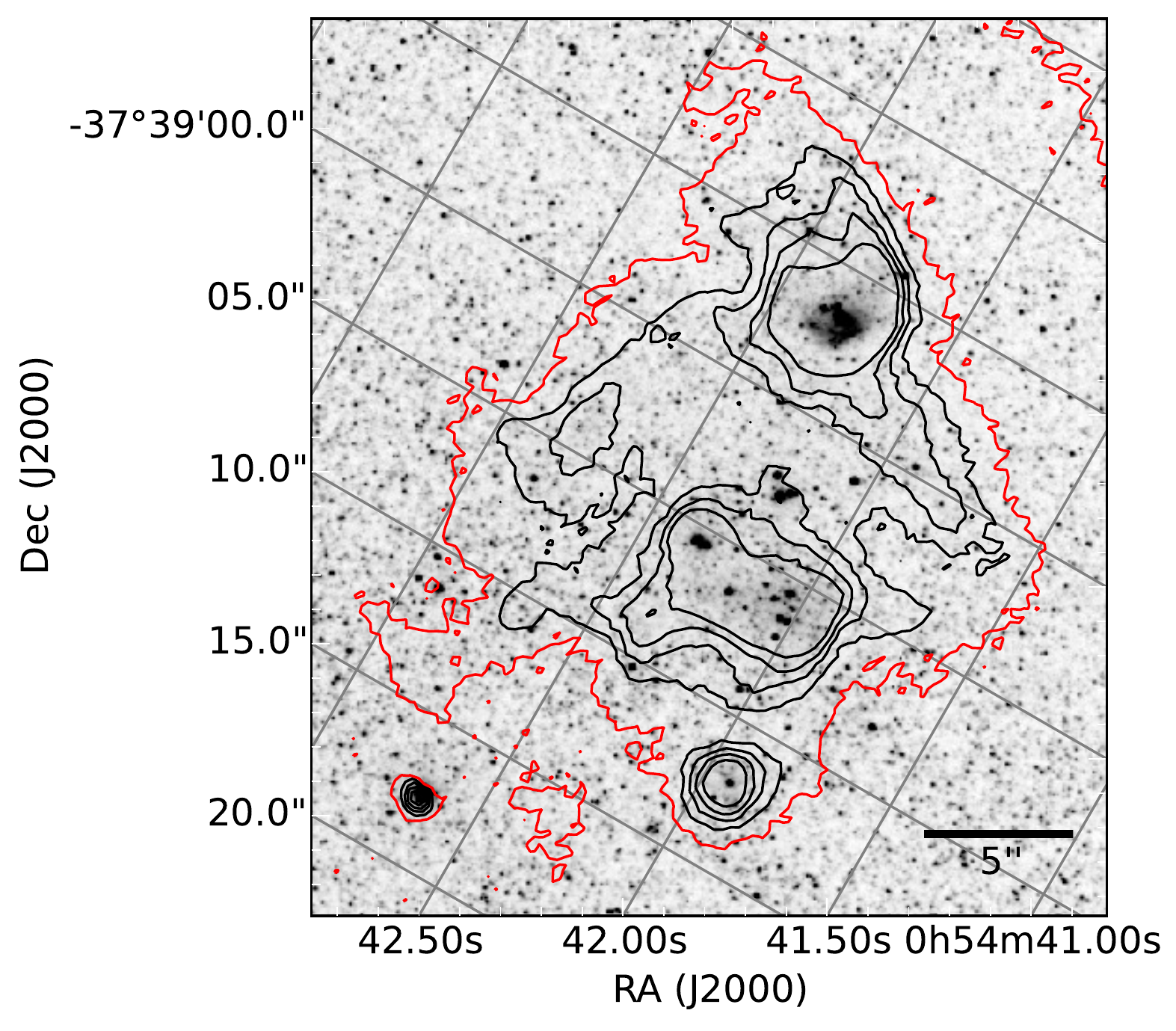}}
  \subfloat[\ion{H}{ii} Region 13]{\label{figur:6_zoom}\includegraphics[width=64mm]{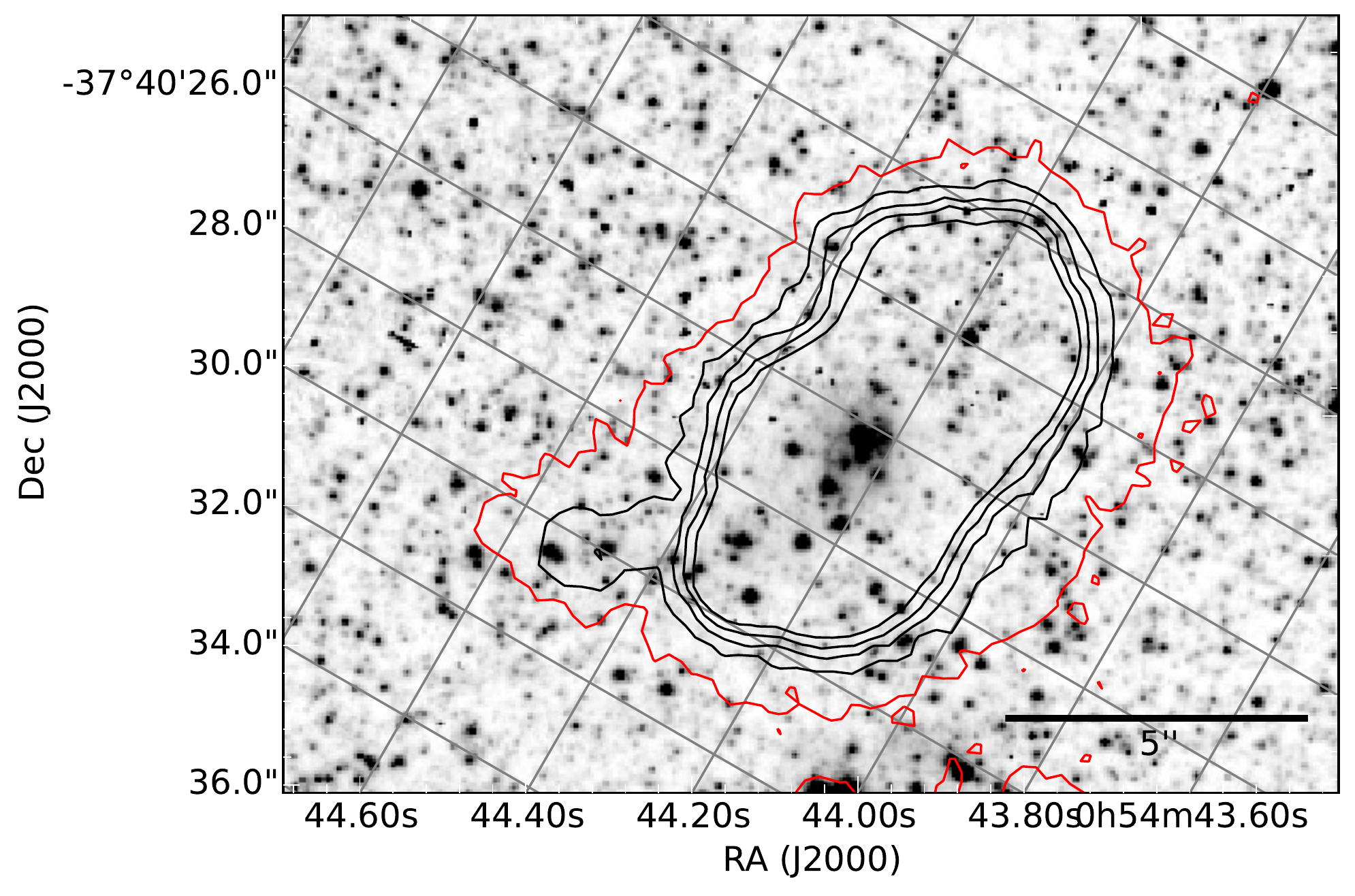}}
  \\
  \subfloat[\ion{H}{ii} region 15]{\label{figur:7_zoom}\includegraphics[width=64mm]{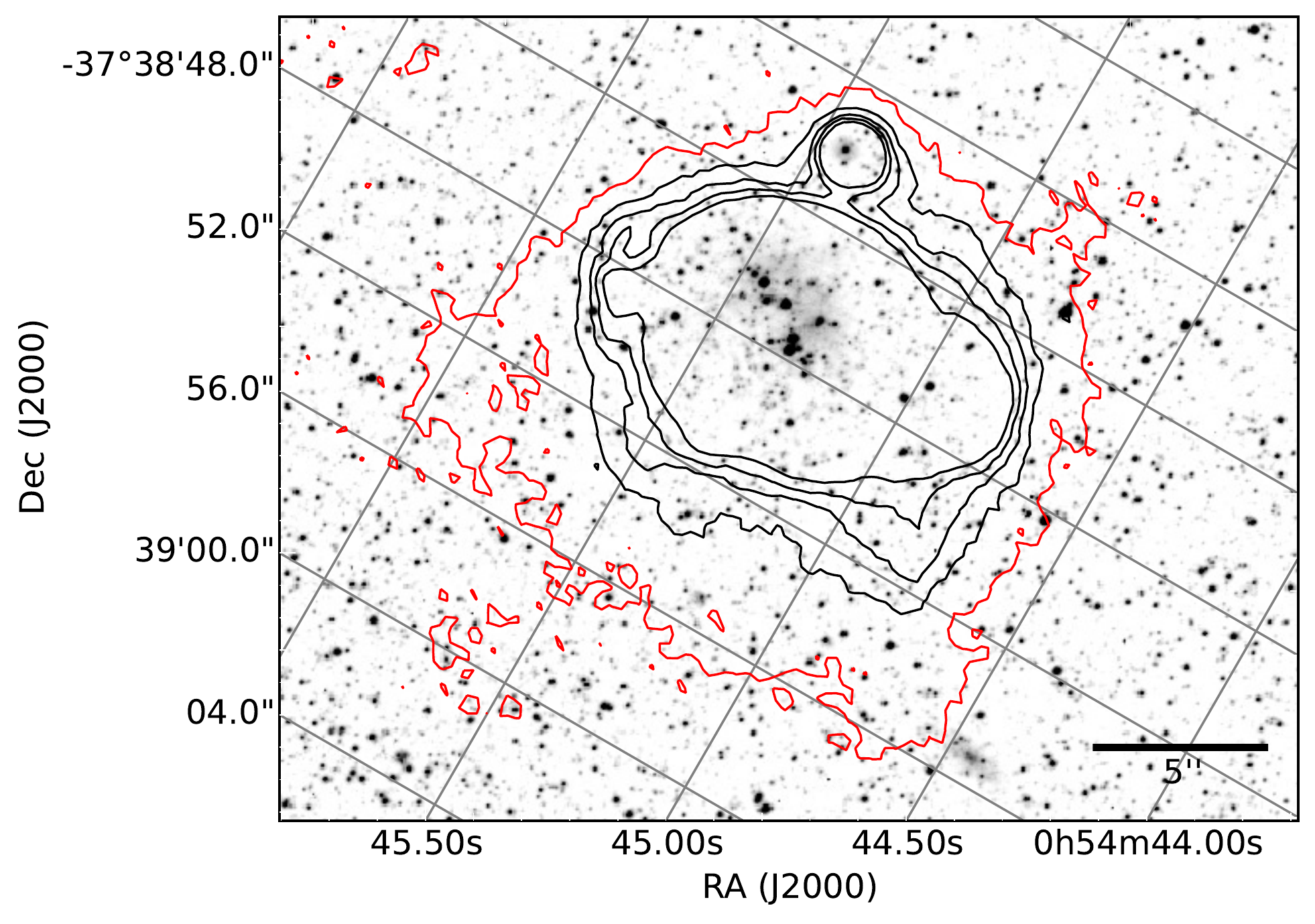}}
  \subfloat[\ion{H}{ii} Region 17]{\label{figur:8_zoom}\includegraphics[width=64mm]{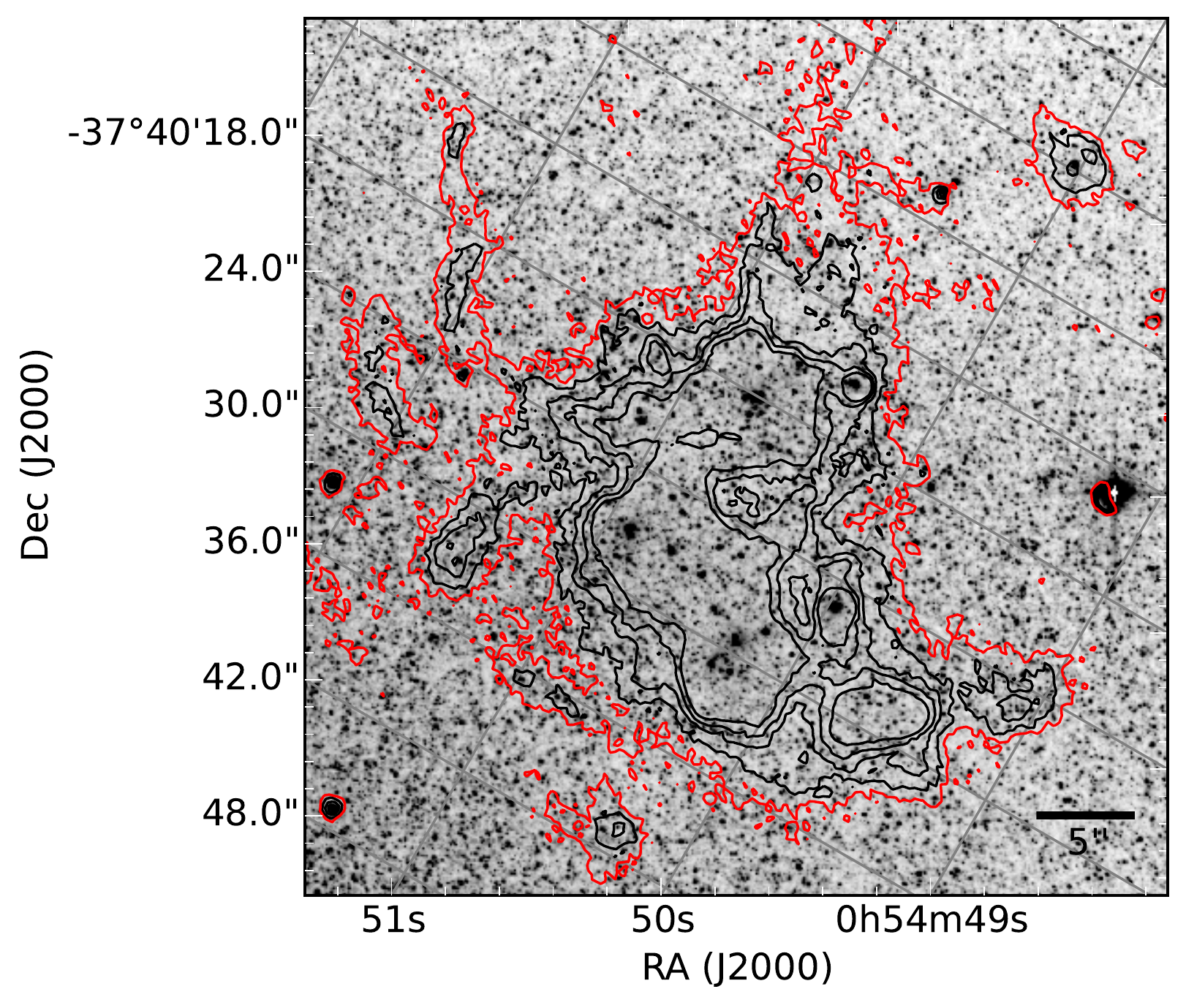}}
  \subfloat[\ion{H}{ii} Region 19]{\label{figur:9_zoom}\includegraphics[width=64mm]{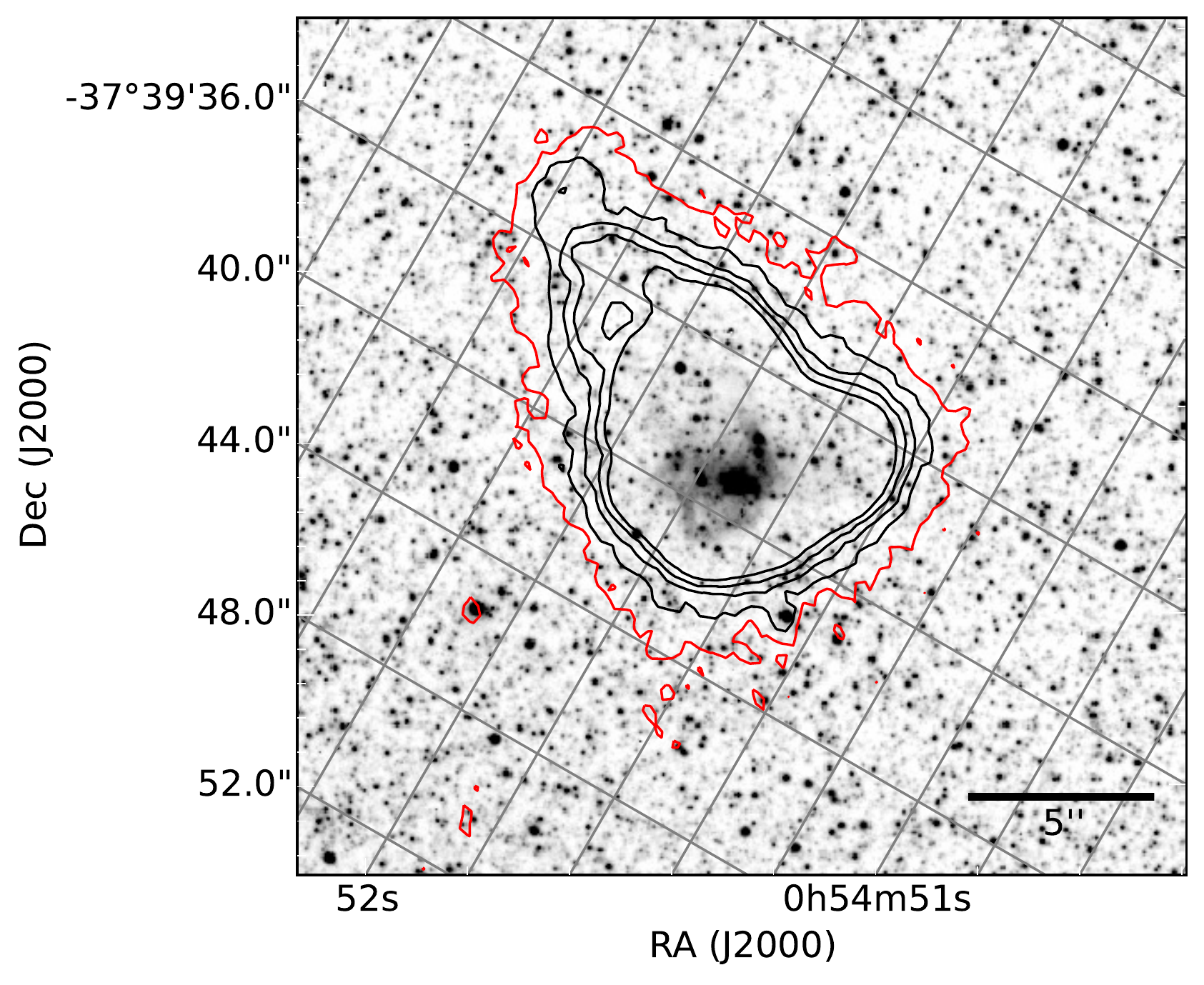}}
  \caption{Detailed view of the shortlisted \ion{H}{ii} regions. The red contours indicate the isophote 4$\sigma$ above the local background inside which the H$\alpha$ flux was measured.}
  \label{fig:HIIRegions_zoom}
\end{figure*}

\begin{table} 
\caption{\ion{H}{ii} region parameters\label{tab:ExcellentRegions}}
\centering
\begin{tabular}{l c c c } 
\hline\hline
\noalign{\smallskip}
ID & $f_{esc}$ & Fitted age & D$_{\mathrm{Center}}$ \\
& (per cent)& (Myr) & (kpc) \\
(1) & (2) & (3) & (4)\\
\noalign{\smallskip}
\hline
\noalign{\smallskip}
2 &  $-850^{+10}_{-3}$  & 8.91 & 3.27 \\
\noalign{\smallskip}
3 &  $-1132^{+15}_{-4}$  & 8.13 & 2.97 \\
\noalign{\smallskip}
4 &  $-483^{+8}_{-3}$  & 6.61 & 3.07 \\
\noalign{\smallskip}
6 &  $3.95^{+0.7}_{-0.1}$  & 5.13 & 3.02 \\
\noalign{\smallskip}
12 & $63.0^{+0.2}_{-0.0}$  & 3.31 & 1.69 \\
\noalign{\smallskip}
13 & $-106^{+17}_{-14}$  & 5.75 & 1.08 \\
\noalign{\smallskip}
15 & $-272^{+61}_{-63}$  & 5.62 & 1.57 \\
\noalign{\smallskip}
17 & $54.1^{+11.4}_{-8.0}$  & 3.55 & 0.48 \\
\noalign{\smallskip}
19 & $22.5^{+13.3}_{-9.0}$  & 3.16 & 0.82 \\

\noalign{\smallskip}
\hline

\end{tabular}
\tablefoot{ Column 1: identification of the \ion{H}{ii} regions, Col. 2: escape fraction in per cent, Col. 3: fitted age of the \ion{H}{ii} region in Myr, Col. 4: distance to the center of the galaxy in kpc.
}
\end{table}

\begin{figure*}[htp]
  \centering
  
  \subfloat[\ion{H}{ii} region 2]{\label{figur:1}\includegraphics[width=64mm]{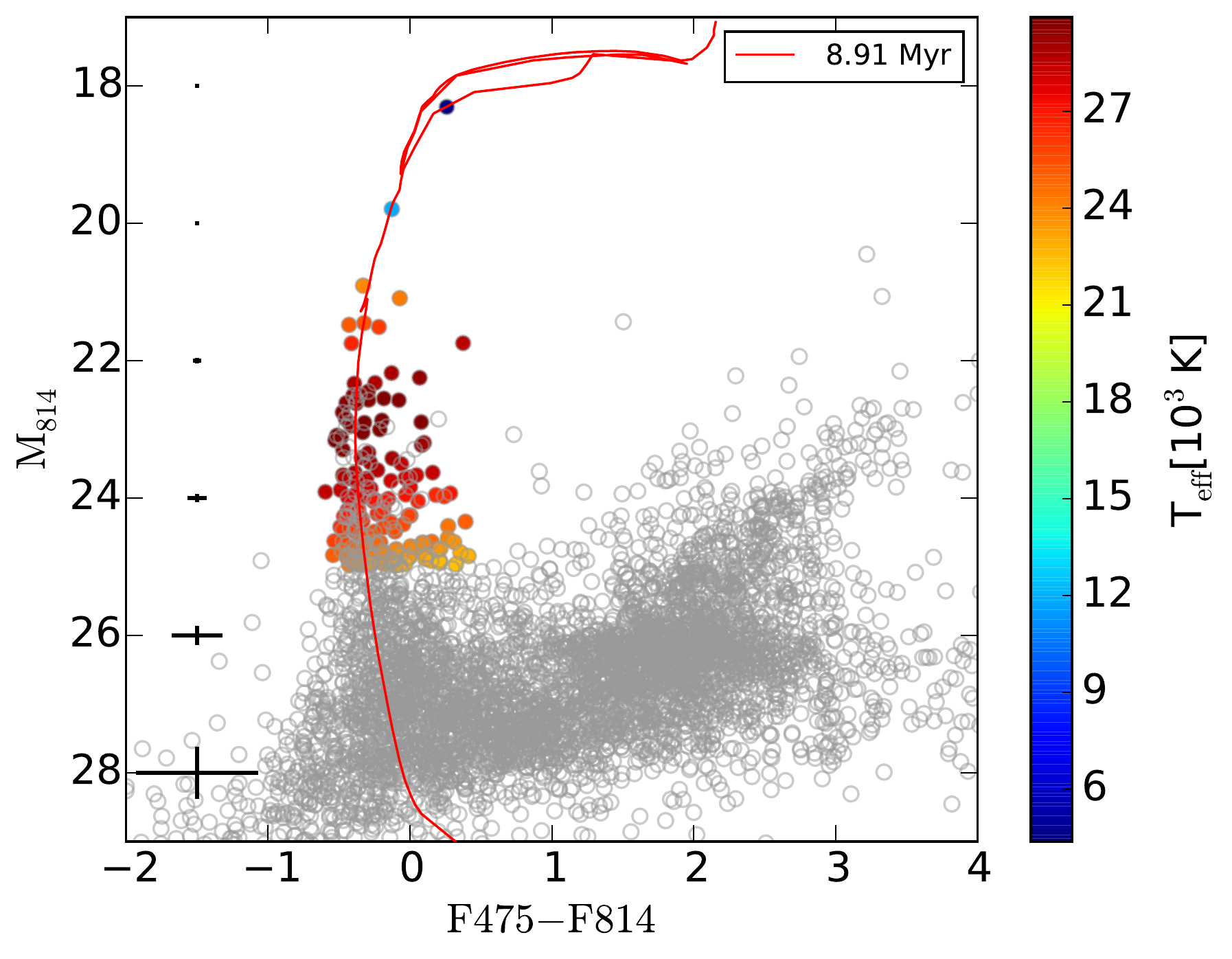}}
  \subfloat[\ion{H}{ii} Region 3]{\label{figur:2}\includegraphics[width=64mm]{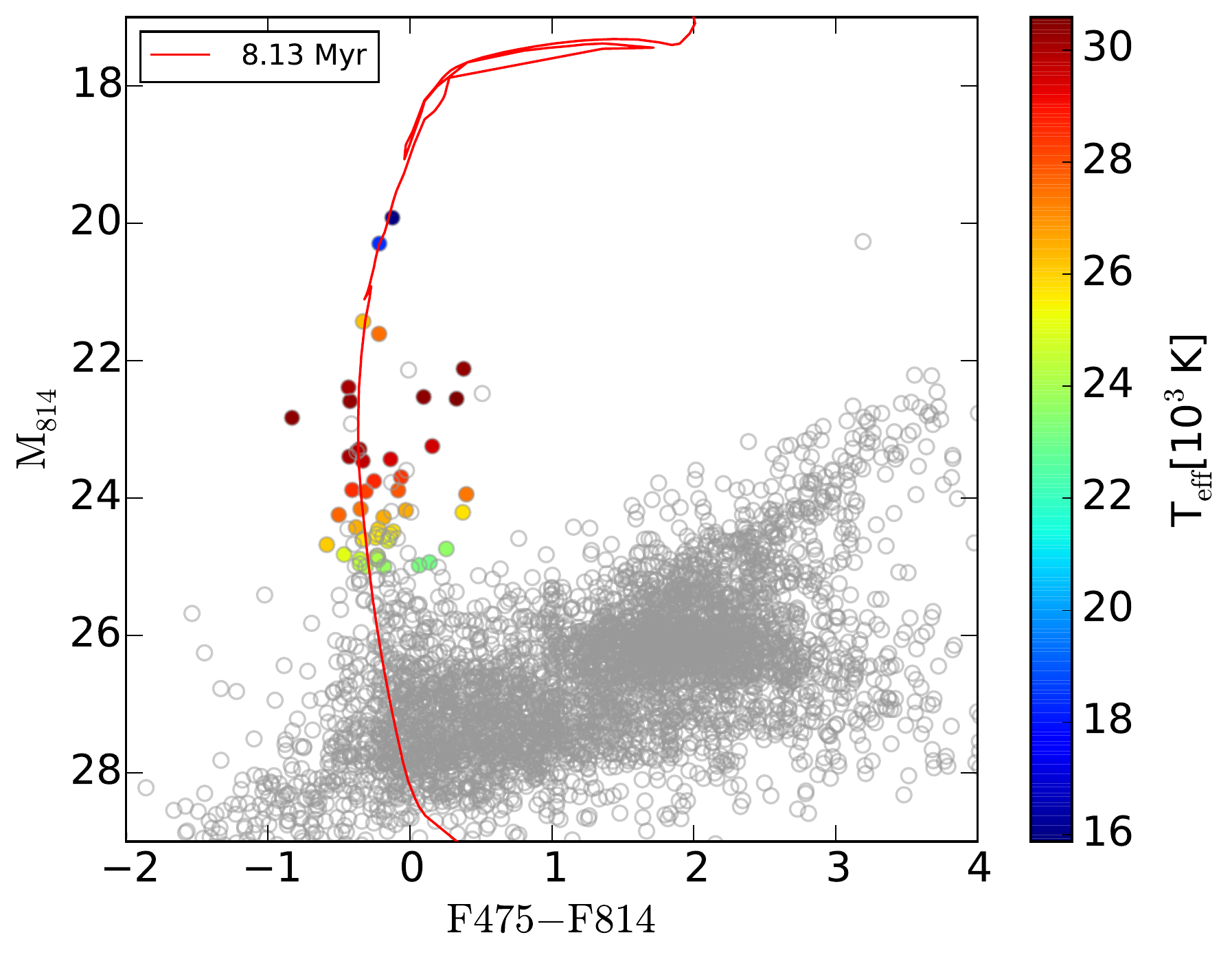}}
  \subfloat[\ion{H}{ii} Region 4]{\label{figur:3}\includegraphics[width=64mm]{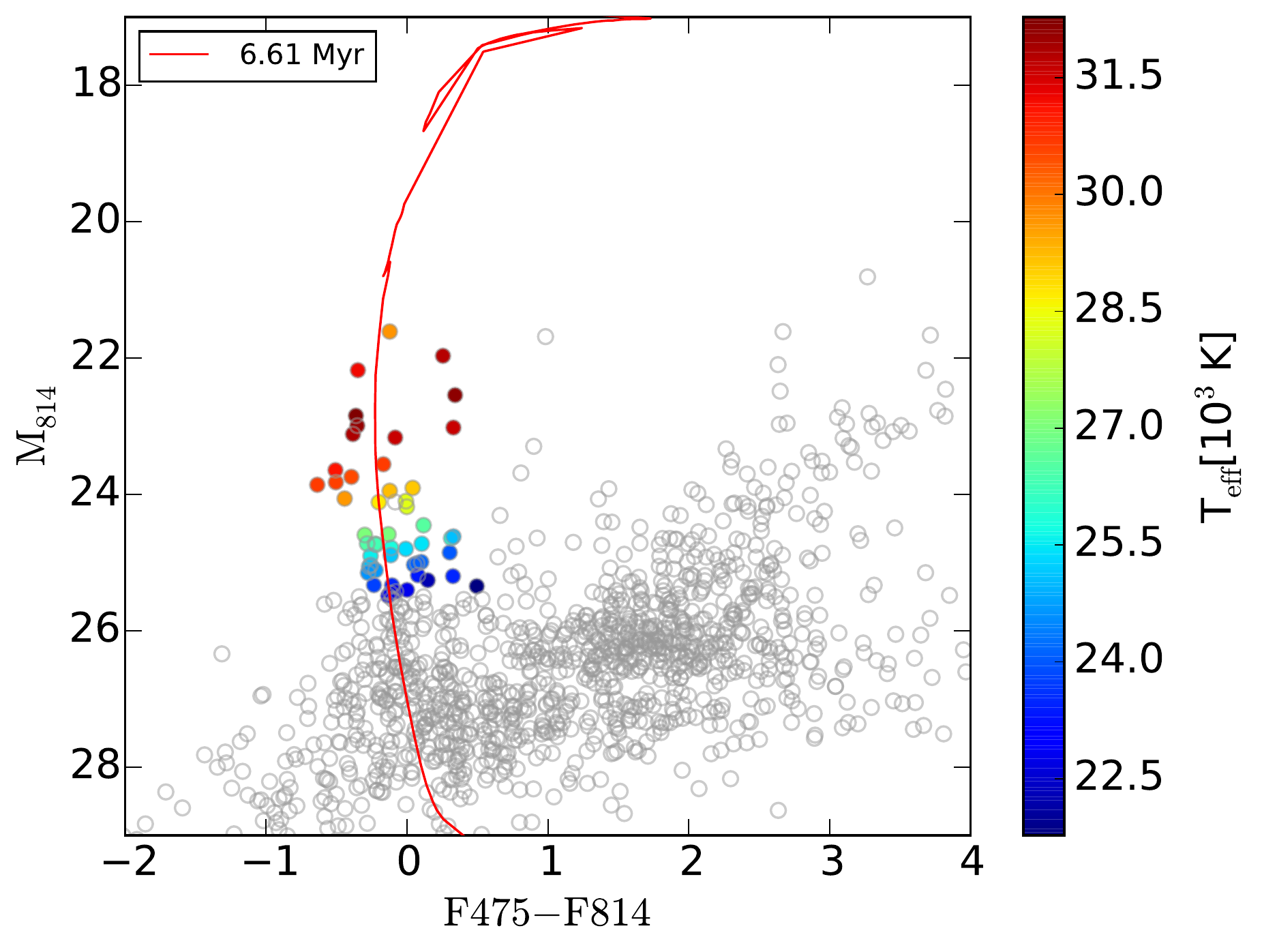}}
  \\
  \subfloat[\ion{H}{ii} region 6]{\label{figur:4}\includegraphics[width=64mm]{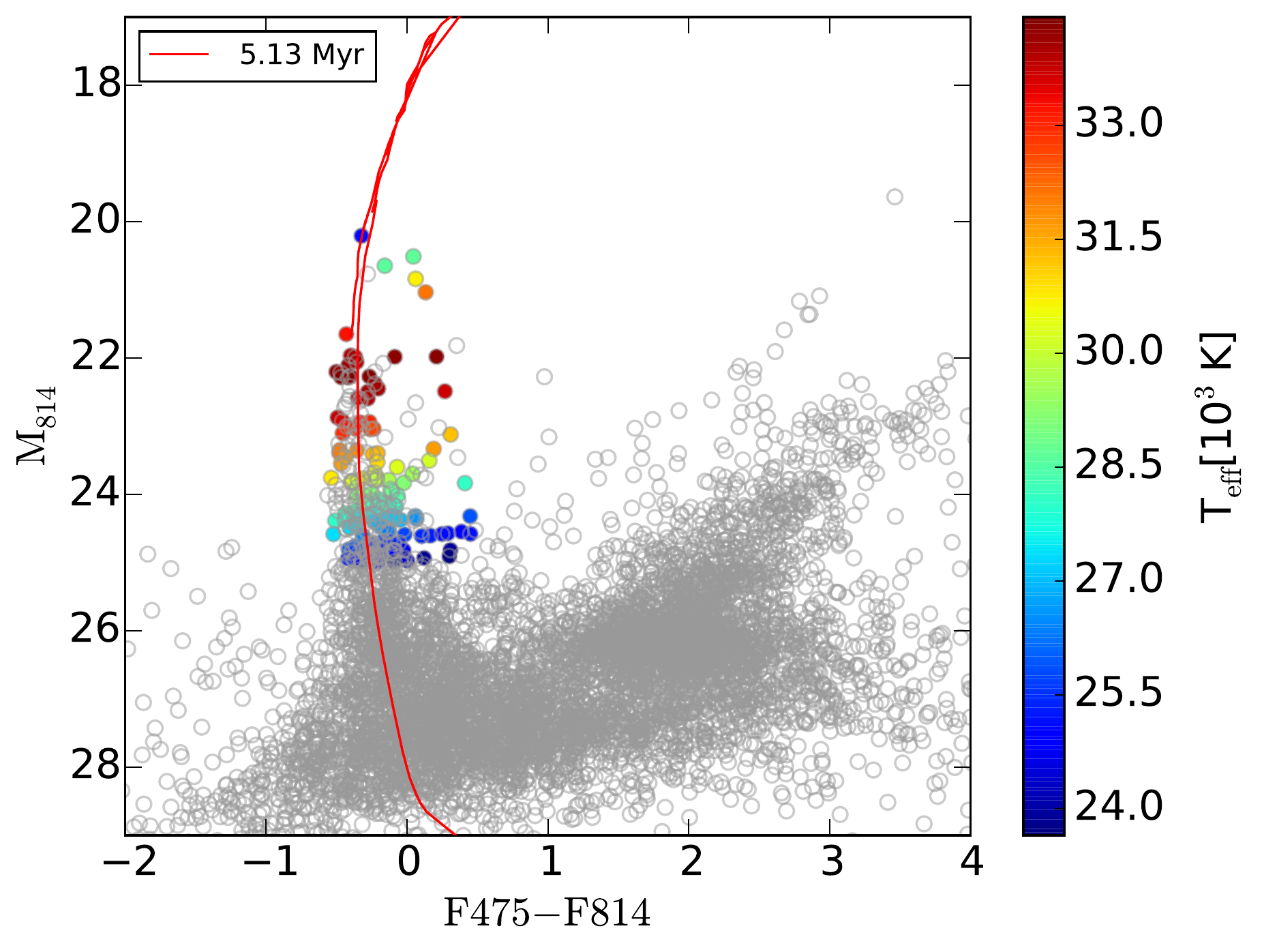}}
  \subfloat[\ion{H}{ii} Region 12]{\label{figur:5}\includegraphics[width=64mm]{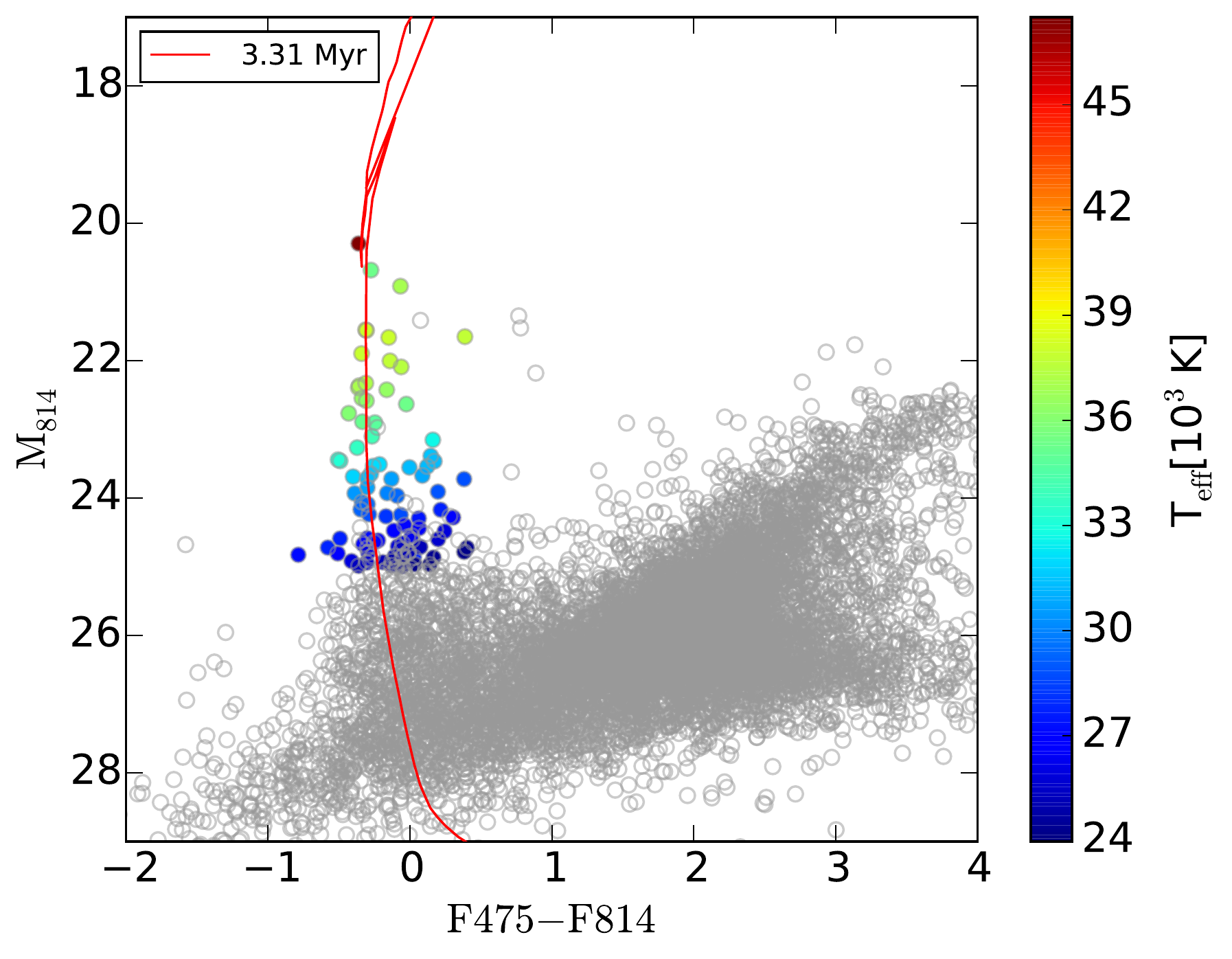}}
  \subfloat[\ion{H}{ii} Region 13]{\label{figur:6}\includegraphics[width=64mm]{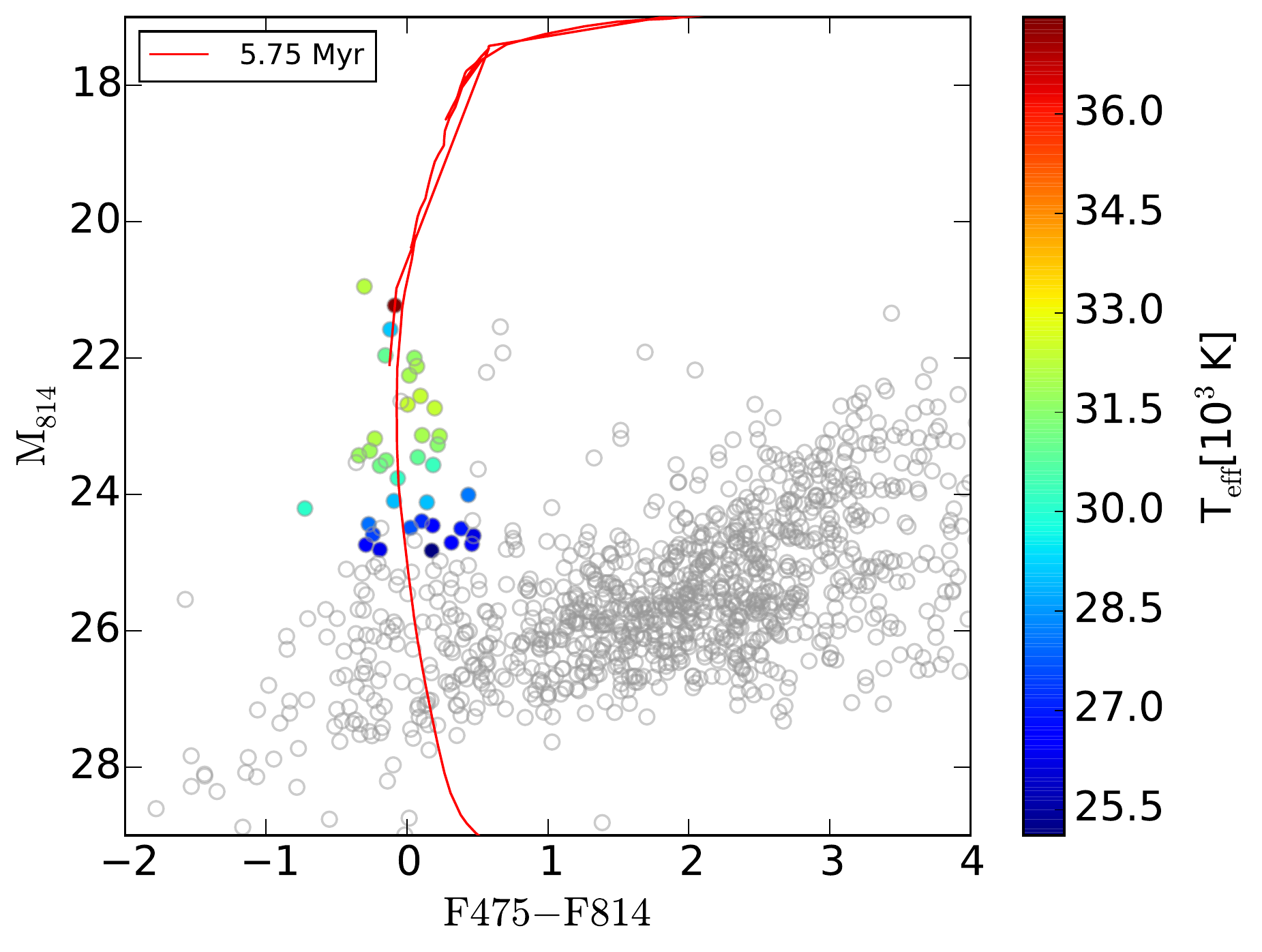}}
  \\
  \subfloat[\ion{H}{ii} region 15]{\label{figur:7}\includegraphics[width=64mm]{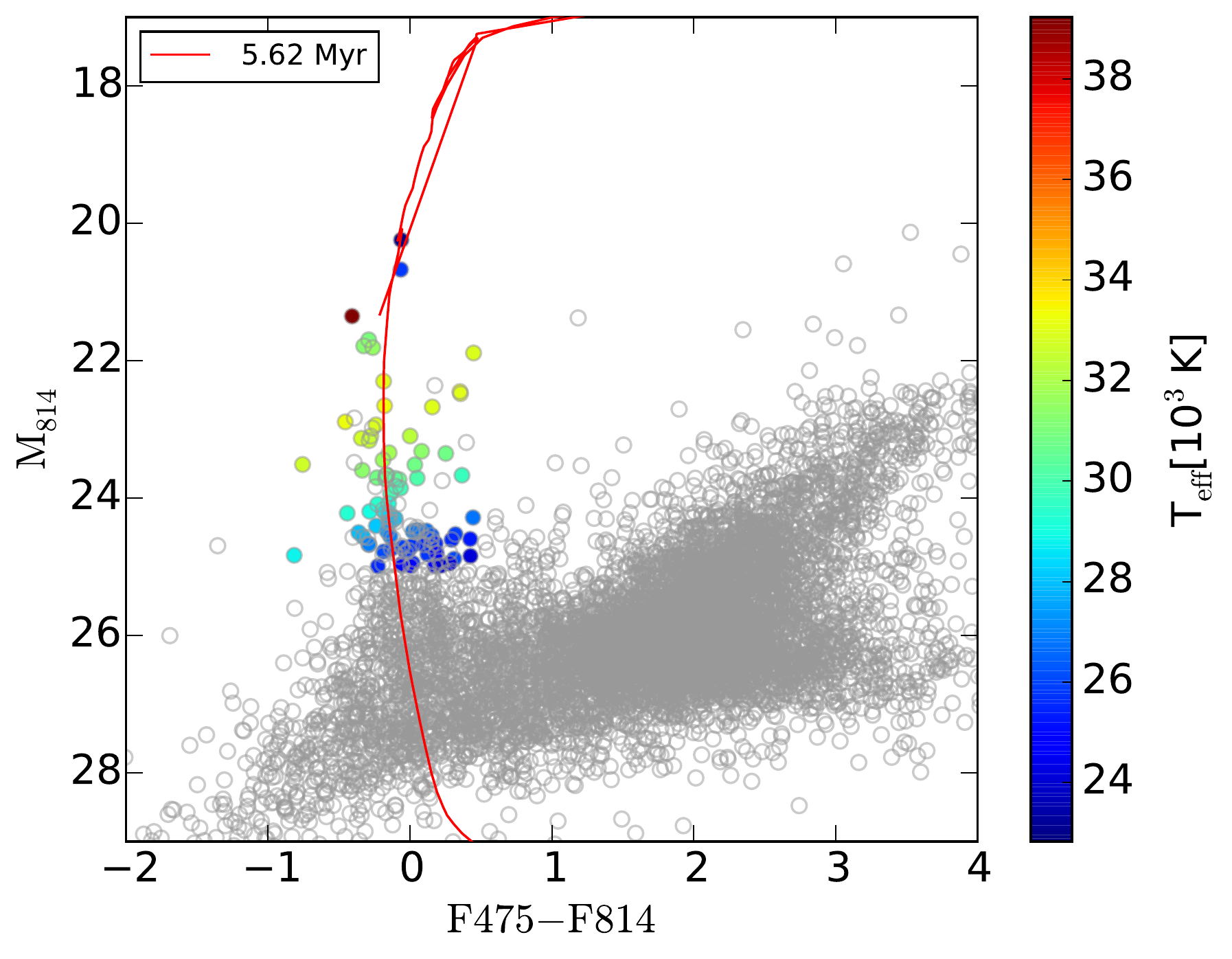}}
  \subfloat[\ion{H}{ii} Region 17]{\label{figur:8}\includegraphics[width=64mm]{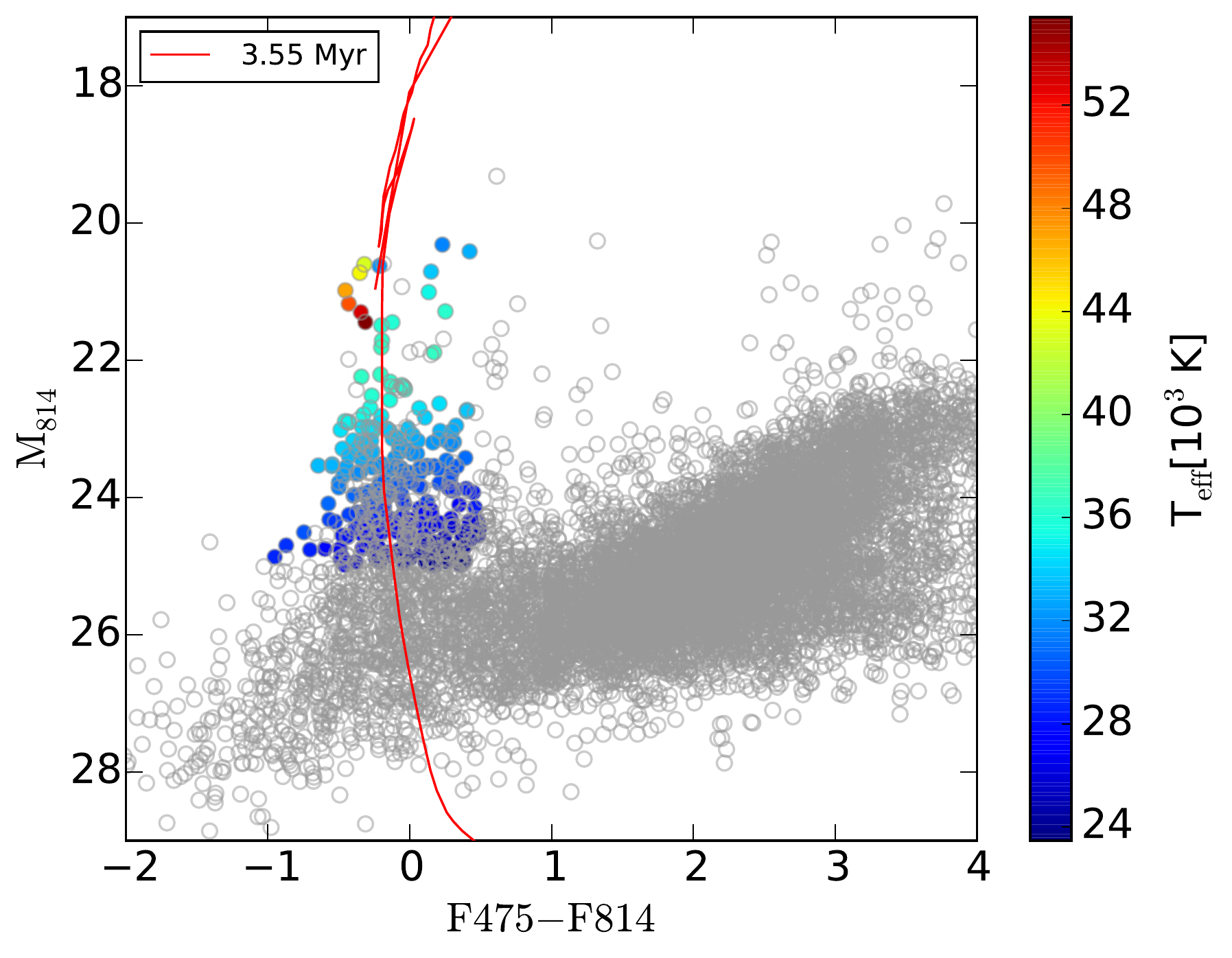}}
  \subfloat[\ion{H}{ii} Region 19]{\label{figur:9}\includegraphics[width=64mm]{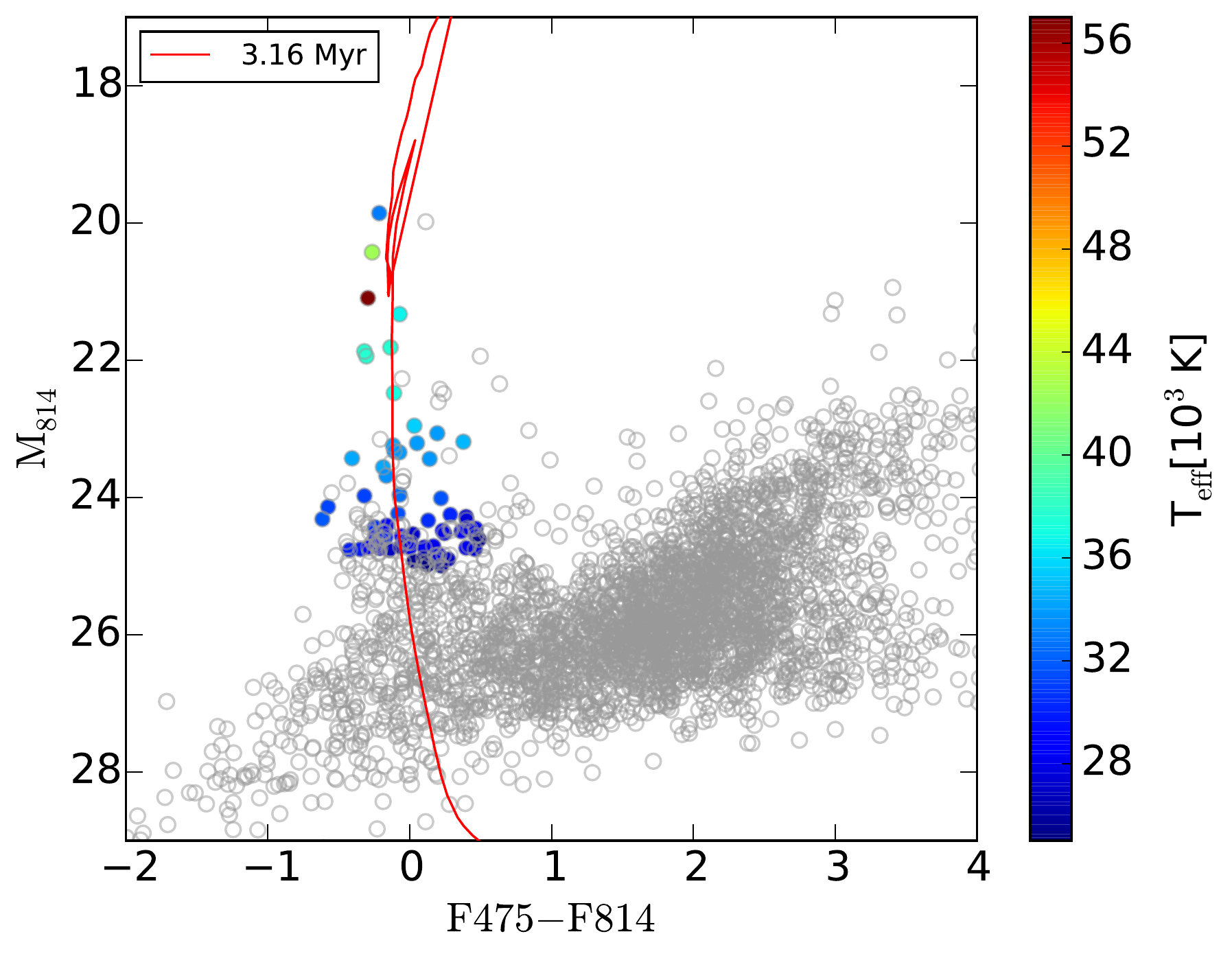}}
  \caption{CMDs of the shortlisted \ion{H}{ii} regions with isochrones that fit the data best. The age of the isochrone is indicated in upper left corner of each panel. The stars for which we fit the parameters using \textsc{ta-da} are color-coded by their fitted temperature. In the upper left panel, the black crosses indicate the typical photometric errors in color and magnitude as a function of the $F814W$ magnitude.}
  \label{fig:CMDs}
\end{figure*}


\subsection{Test of the method}
To check how reliable our method of fitting stellar parameters is, we carried out additional tests with artificial star clusters. We also checked whether we would benefit from adding a supplementary filter and performed the tests using $B$,$V$ and $I$ photometry on the one hand, which corresponds approximately to the used $HST$ filter combination, and on the other hand using the $U,B,V$ and $I$ photometry because the $U$ filter is more sensitive to the emission peak of hot stars. We created synthetic photometry of 19 stellar clusters based on the theoretical Padova isochrone set \citep{Marigo08}. The clusters contain a random number of stars between 30 and 120. The stars themselves were chosen to have masses of between 15 and 100~M$_{\sun}$ with an underlying Salpeter logarithmic initial mass function (IMF) with an index $\alpha$ of $-$2.35 \citep{Salpeter55}. Extrapolated to a lower mass of 1~M$_{\sun}$ using the same IMF, the clusters have masses of between 3.7$\times10^3$ and 1.1$\times10^4$~M$_{\sun}$. Each cluster has a random mean age of between 4 and 10~Myr with an allowed Gaussian scatter between 0.1 and 0.5~Myr around it. To the stellar magnitudes of each cluster we added a Gaussian noise of 0.01 mag as well as an extinction value that consists of an mean foreground extinction plus an additional lognormal scatter.  

\begin{figure}
\centering
\begin{tabular}{l}
\includegraphics[width=8cm]{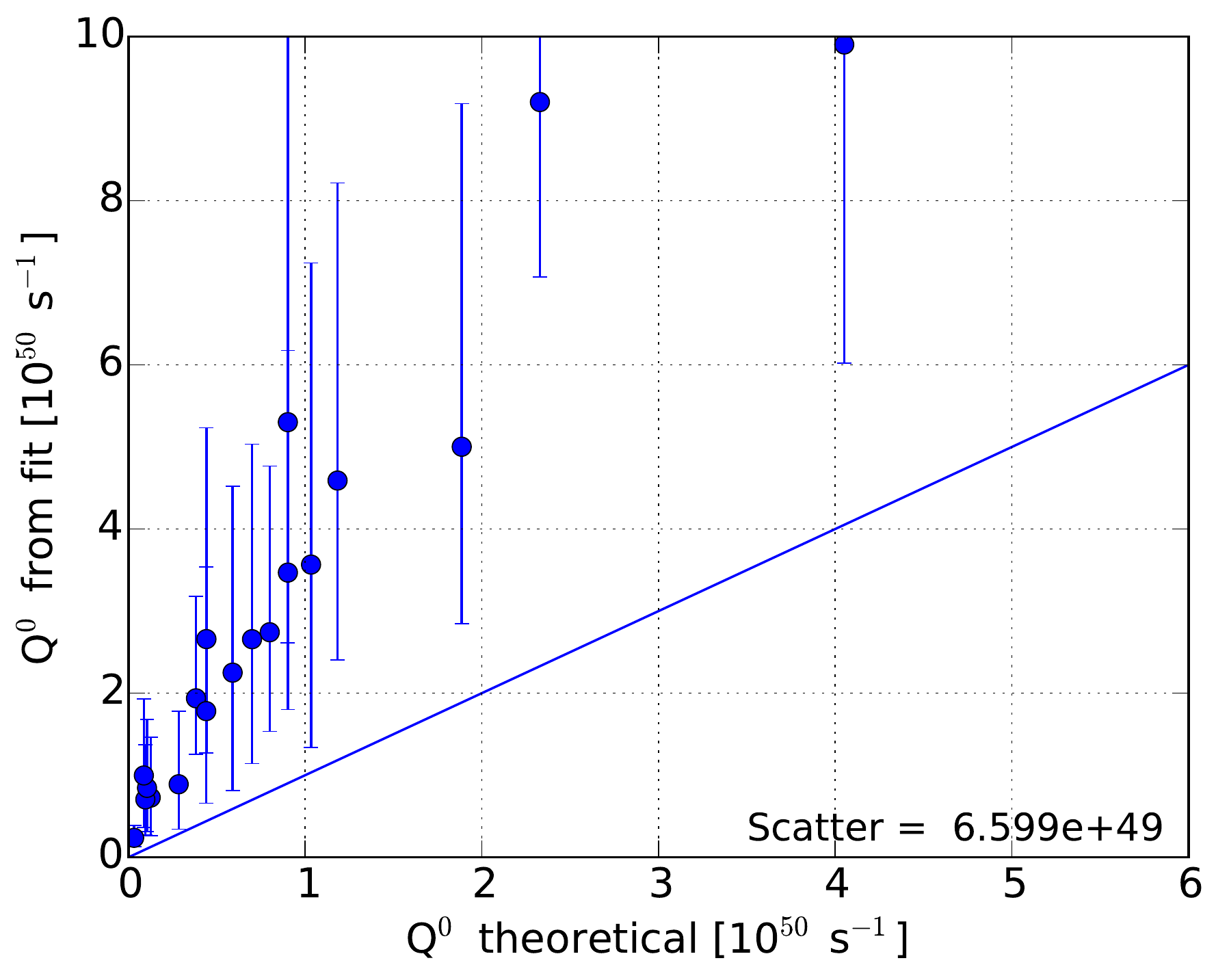} \\
\includegraphics[width=8cm]{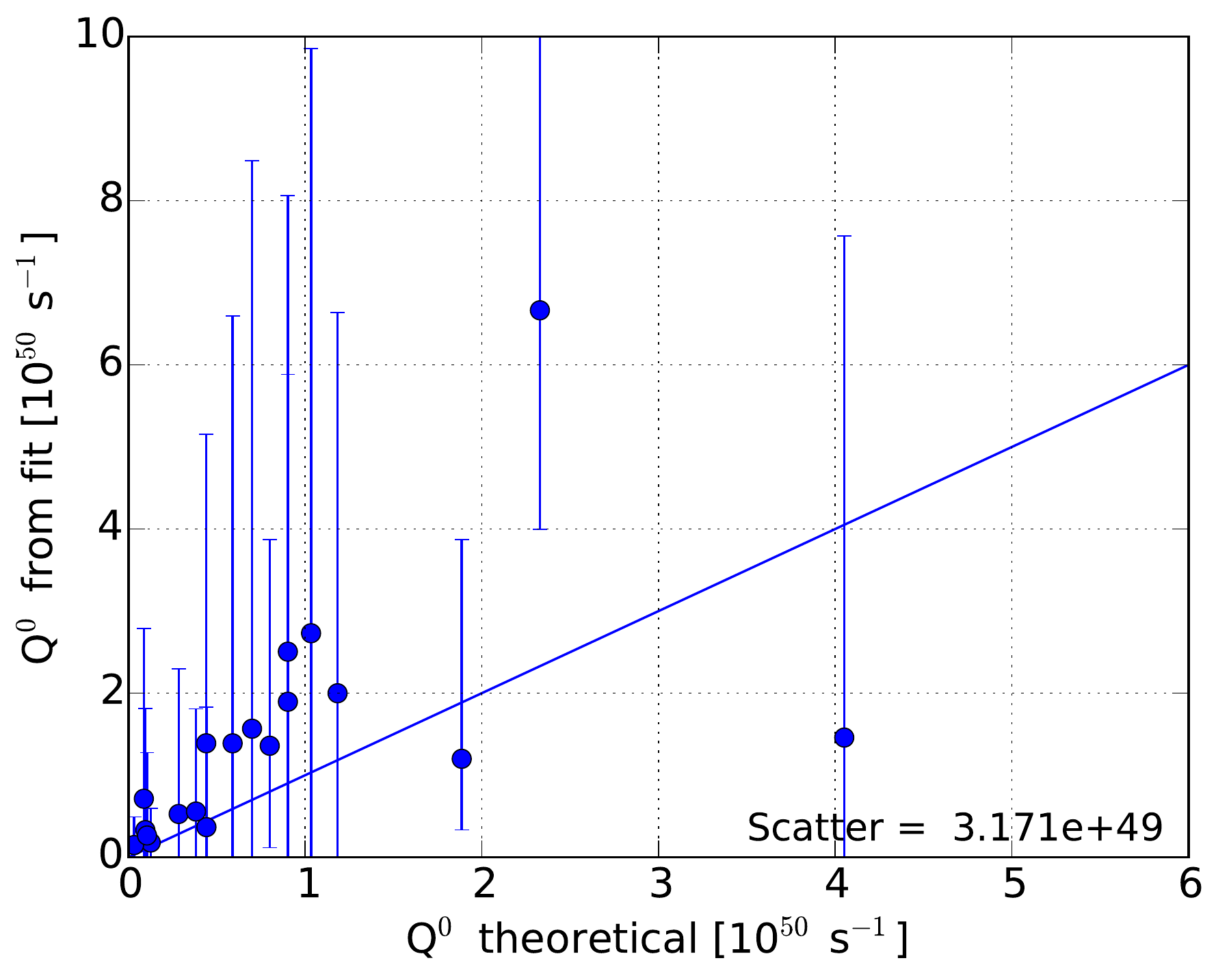} 
 \end{tabular}
\caption{Fitted ionizing photon flux as a function of the real ionizing photon emission rate without any a priori assumptions about the ages and extinction. The blue solid line is the one-to-one relation. \textit{Top}: for $UBVI$ photometry, \textit{bottom}: for $BVI$ photometry.}
\label{fig:Q0_allvsFit}
\end{figure}

\begin{figure}
\centering
\begin{tabular}{l}
\includegraphics[width=8cm]{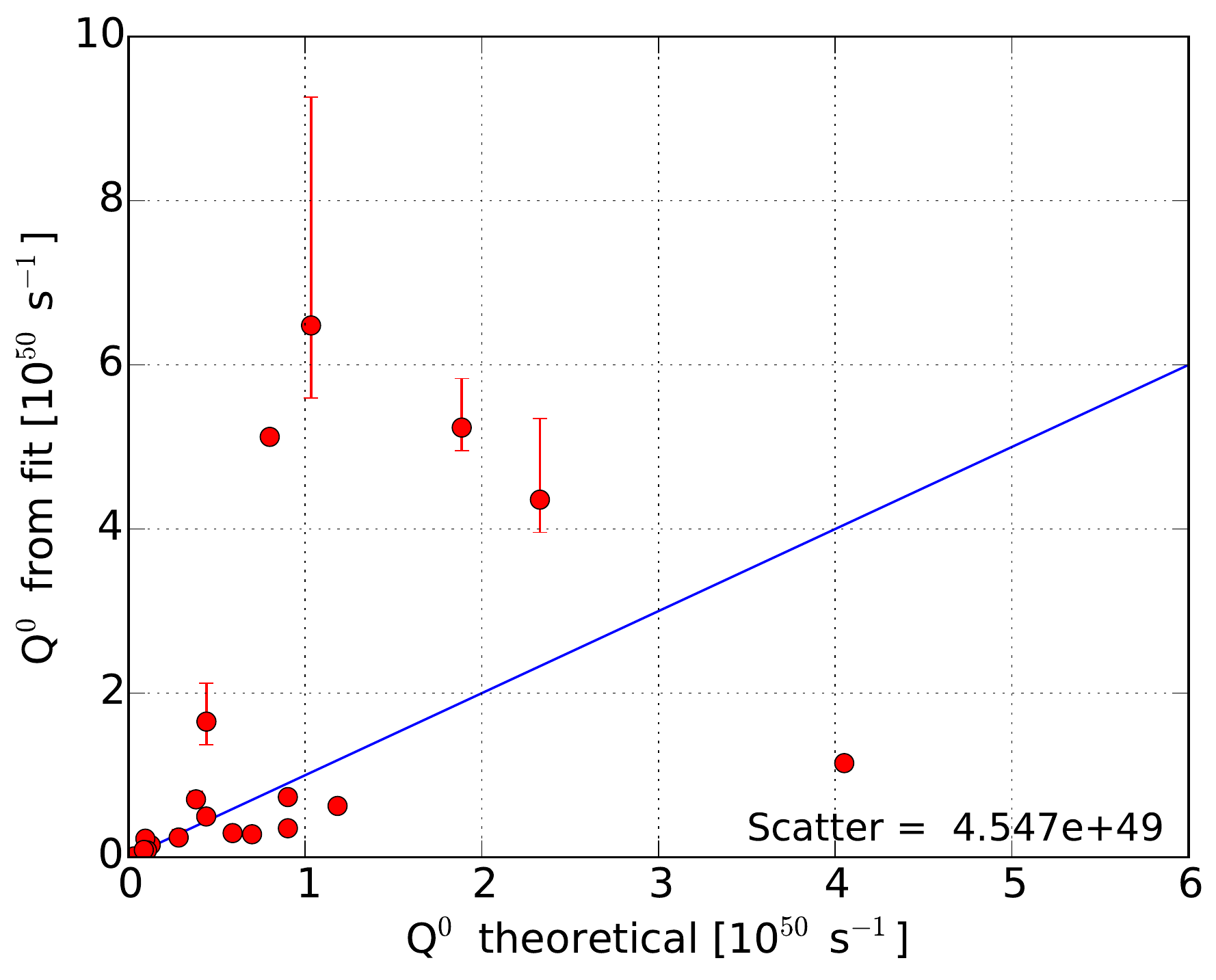} \\
\includegraphics[width=8cm]{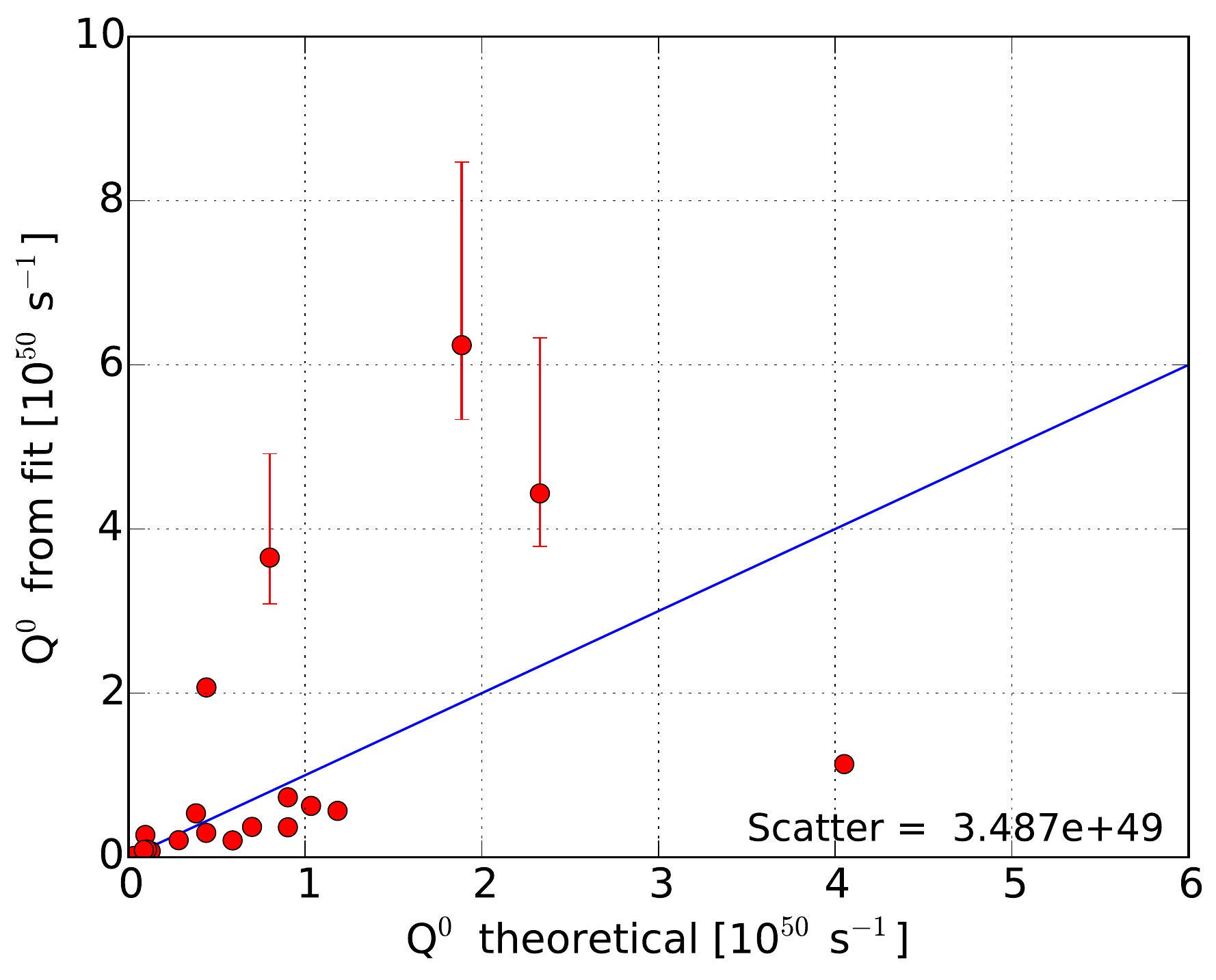} 
 \end{tabular}
\caption{Fitted ionizing photon flux as a function of the real ionizing photon emission rate. Before fitting with \textsc{ta-da,} we fit a single isochrone to the CMD to constrain the age and extinction of the synthetic cluster. The blue solid line is the one-to-one relation. \textit{Top}: for $UBVI$ photometry; \textit{bottom}: for $BVI$ photometry.}
\label{fig:Q0_minvsFit}
\end{figure}

\begin{figure}
\centering
\begin{tabular}{l}
\includegraphics[width=8cm]{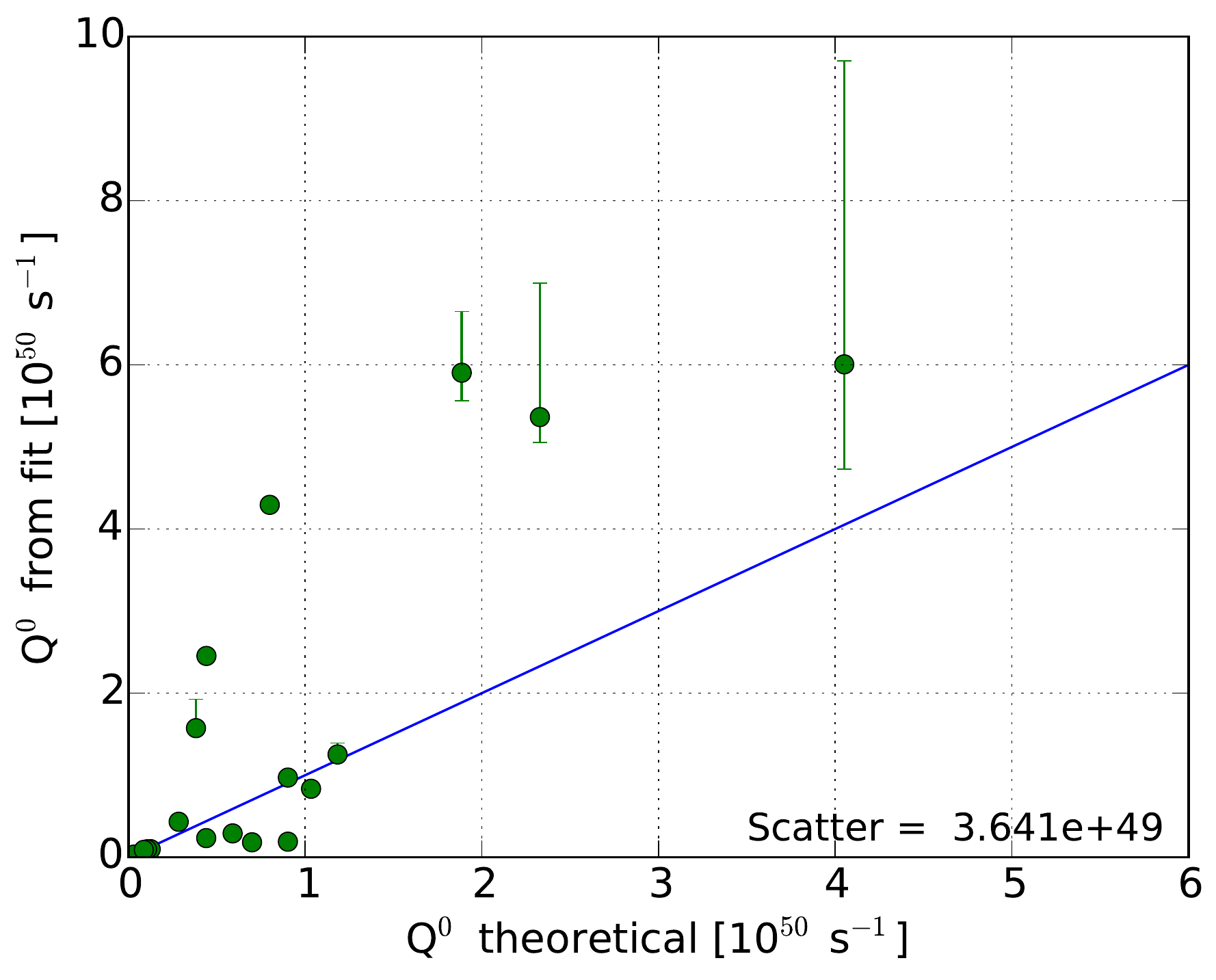} \\
\includegraphics[width=8cm]{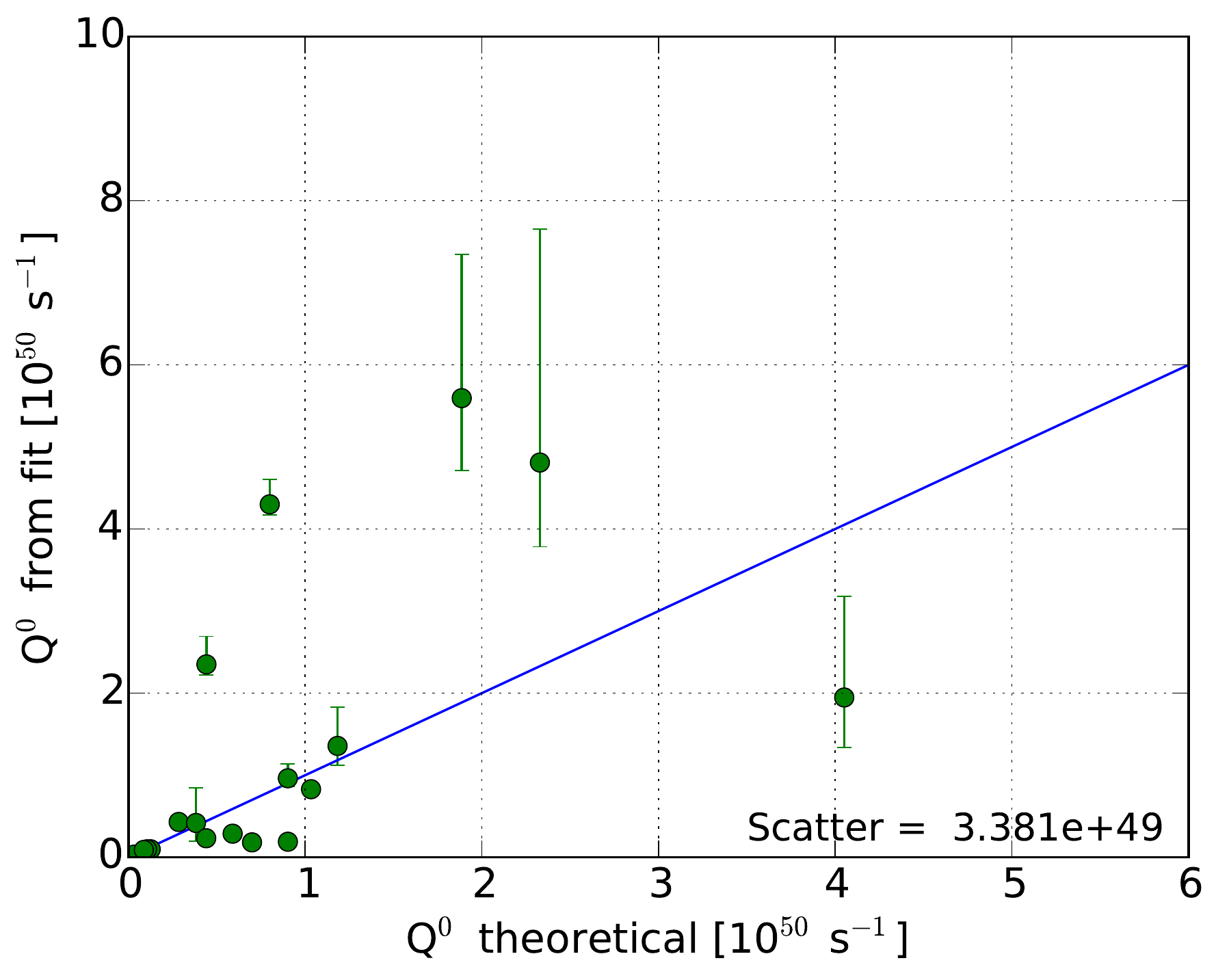} 
 \end{tabular}
\caption{Fitted ionizing photon flux as a function of the real ionizing photon emission rate using the mean real ages and extinctions of the clusters. The blue solid line is the one-to-one relation. \textit{Top}: for $UBVI$ photometry; \textit{bottom}: for $BVI$ photometry.}
\label{fig:Q0_meanvsFit}
\end{figure}

We fit the parameters of the created stellar associations in three different ways:
\begin{enumerate}
\item Without any a priori constraints on the age of the stars and with an extinction $A_V$ ranging from 0.0 to 0.7~mag, so
that we cover the range of mean extinction values of our created clusters and to allow higher values as well.
\item Constraining the stellar ages and extinction values to the real mean values of the cluster.
\item Fitting of theoretical isochrones to the CMDs of the clusters, as we did for the \ion{H}{ii} regions of NGC~300.
\end{enumerate}
After the fitting we computed the ionizing photon emission rates of the synthetic clusters for each of the three different fits and compared it to the true value using the BT-Settl2010 model atmosphere grids to test the reliability of the three different methods. 
Fitting without a priori assumptions is a highly degenerate problem and the results are therefore very uncertain, both for the $UBVI$ and the $BVI$ filter set (see Fig.~\ref{fig:Q0_allvsFit}). However, for the $UBVI$ filter set, the results show a systematic offset.
Even if we set our fitting tool with pre-defined age and extinction values, the final results show clear discrepancies between the fit and true $Q^0$ values for both the mean real ages and extinctions and those determined by isochrone fitting to the CMD (see Figs.~\ref{fig:Q0_minvsFit} and \ref{fig:Q0_meanvsFit}). The error bars in all figures represent the maximum errors resulting from the uncertainties in the fitted stellar temperatures as given by \textsc{ta-da}.
For most of the clusters, the parameters of the fitted isochrone agree well with the real mean ones, which explains the comparable scatter around the one-to-one relation for both methods. For the $UBVI$ ($BVI$) filter photometry, 14 (15) out of 19 clusters have fitted ages that are within $\pm$15\% of the real mean value. The fitting of the mean extinction was less accurate. For the $UBVI$ ($BVI$) filter combination, only 6 (7) clusters are within $\pm$15\% of the real mean value and 12 (13) within $\pm$30\%. Surprisingly, the usage of the additional U filter does not improve the result of $Q^0$ for all of the three methods (see Fig.~\ref{fig:Q0_minvsFit}). 

\begin{figure}
\centering
\includegraphics[width=7cm]{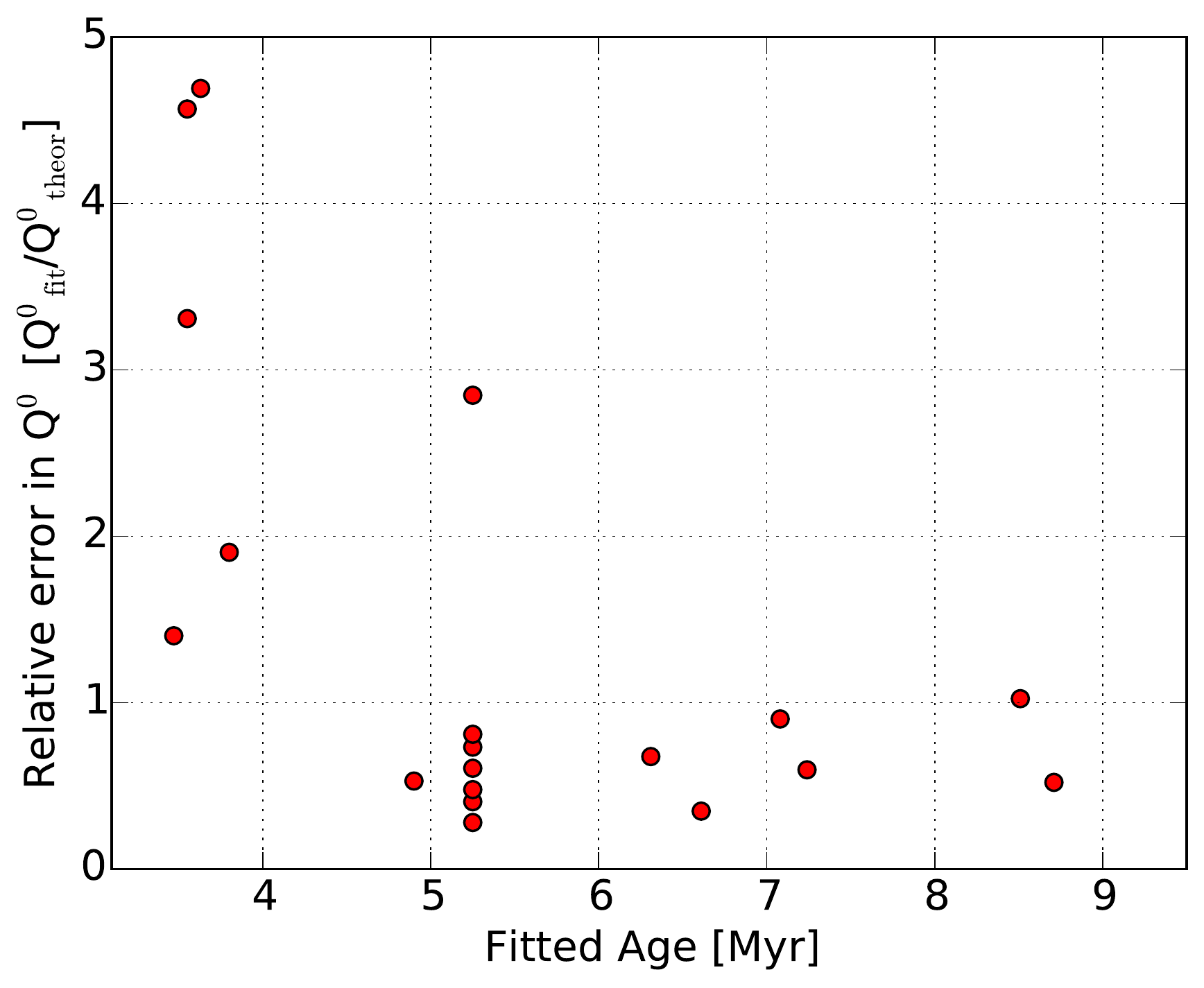}  
\caption{Relative error in the fitted value of $Q^0$ as a function of the fitted age of the cluster using $BVI$ photometry.}. 
\label{fig:AgevsQ0error}
\end{figure}

\begin{figure}
\centering
\includegraphics[width=7.0cm]{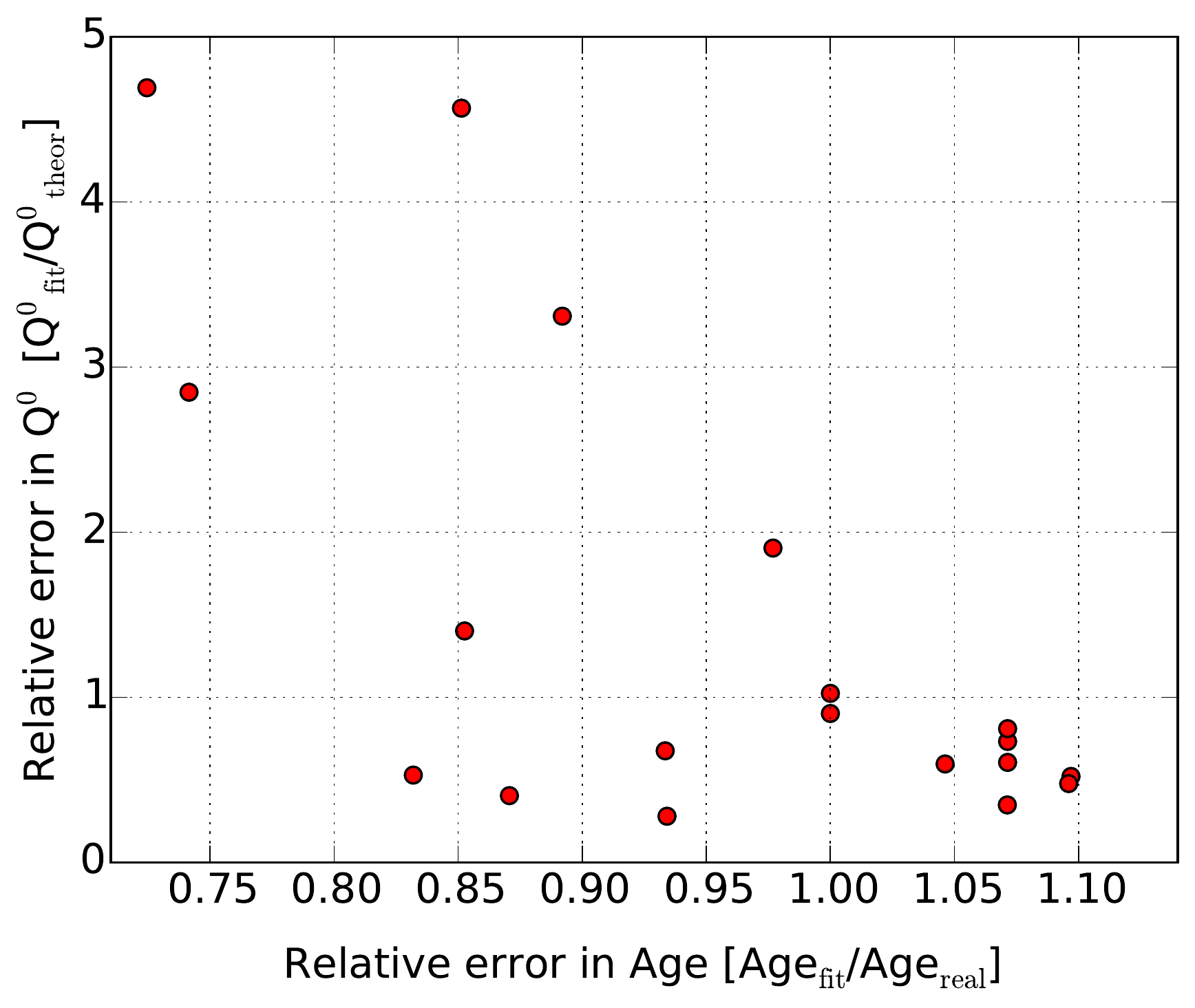} 
\caption{Relative error of $Q^0$ as a function of the relative error of the fitted cluster age using $BVI$ photometry.}
\label{fig:Age_error_vs_q0_error}
\end{figure}

\begin{figure}
\centering
\includegraphics[width=7.5cm]{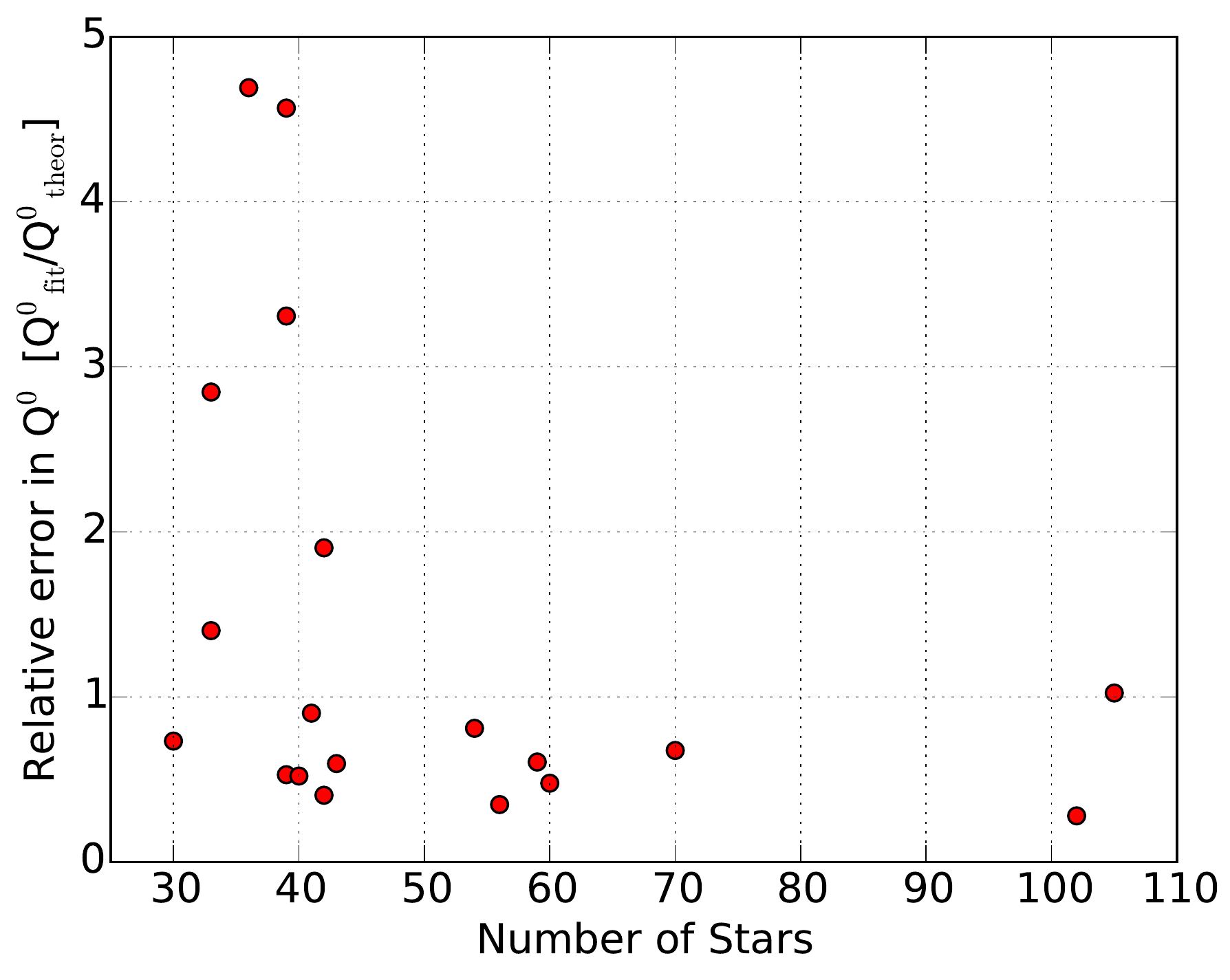} 
\caption{Relative error of $Q^0$ as a function of the number of stars in the cluster using $BVI$ photometry.}. 
\label{fig:NstarsvsQ0error}
\end{figure}

The results of the fitting might depend on the cluster properties. Therefore,  we correlated the relative errors of the ionizing photon flux with the parameters of the test clusters and searched for any trends.  Figure~\ref{fig:AgevsQ0error} shows the relative error of the fitted $Q^0$ value as a function of the fitted age of the cluster. For young ages ($\lesssim$5~Myr) we are more likely to overestimate the ionizing photon output of the cluster, while for older fitted ages we obtain $Q^0$ that is lower than the real values. This is also evident in Fig.~\ref{fig:Age_error_vs_q0_error}, where we plot the relative error in $Q^0$ as a function of the relative error in the fitted age of the cluster. The greater
the number of hot stars in the cluster, the smaller the scatter in the relative error of $Q^0$ (Fig.~\ref{fig:NstarsvsQ0error}). This is as expected because the greater the number of stars, the smaller the contribution of a single star. In clusters with fewer than 50 stars, the relative error in $Q^0$ is between 0.4 and 4.8. We note, however, that in clusters with more than 50 stars we systematically underestimate the ionizing photon flux. There, our fitting yields between 0.3 and 1.0 times the real value of $Q^0$.

\section{Discussion and conclusions\label{sec:disc}}

We examined the feasibility of inferring the escape fraction of LyC photons from individual \ion{H}{ii} regions within the nearby spiral galaxy NGC~300 using broadband and narrowband photometric data. Our approach was to determine the integrated $Q^0$ value of the stars within the young clusters from broadband photometry and stellar model atmospheres. However, we have shown here that the uncertainties inherent to spectral typing using broadband data completely dominate the results. Our tests using artificial clusters showed that we are unable to correctly reconstruct the stellar parameters from photometric data in three or four bands even under idealized conditions. The tests assumed that the young clusters consist of a coeval population, which may not necessarily be true. There might also be a variable amount of reddening throughout the region because the young clusters are probably located within a patchy cloud.
We accounted for these facts and allowed a small scatter in age and extinction within the clusters in our tests. As a result
of these uncertainties, stars are fit to an incorrect isochrone, leading to incorrect stellar parameters. As some regions are also sparsely sampled in the upper parts of the CMD, an accurate fitting of the mean age and extinction is not possible, which
increases the errors in the integrated $Q^0$ even more. In our analysis and tests, we ignored other factors that also influence the determination of the stellar spectral types, such as photometric crowding and unresolved binary stars.

In our tests with artificial star clusters we found that we tend to underestimate the LyC luminosity for fitted ages older than about 5~Myr. This is reflected in the results of $f_{esc}$ of the real \ion{H}{ii} regions. All regions fit to be older than about 5.5~Myr have "negative" escape fractions, confirming that the inferred $Q^0$ values are too low. All regions with younger ages are found to leak ionizing photons with escape fractions of up to $\sim$66\%. The reason is that at younger ages the $Q^0$ is generally overestimated by large factors. This apparent trend is against the expectations, as the escape fraction should increase with increasing age of the cluster (e.g., \citealt{Dale13}).

In our simplified analysis we have assumed that all stars within a single \ion{H}{ii} region have the same age and are equally affected by extinction. However, this is not necessarily true
because most of the clusters in our sample show a broadened main sequence or even a second sequence at redder colors, which disagrees with photometric errors (see Fig.~\ref{fig:CMDs}).  This broadening can be caused by stellar blending or crowding, unresolved stellar binaries, or differential extinction, or a combination of several effects. To explain the stars that are scattered to the red side of the main sequence, age spreads of about 30~Myr or more must be assumed, which is unreasonable. Therefore age spreads seem to be an unlikely explanation. From our data set we are not able to tell which other effects might cause this broadening. If a varying reddening throughout the \ion{H}{ii} region is responsible for it, then a difference in extinction $\Delta \mathrm{A_{V}}$ of up to 1.25~mag is required to explain the spread. If we allow a varying extinction to cover the width of the main sequence, then our estimates of the stellar temperatures and surface gravities, and thus of $Q^0$ , are very different from the previous ones with a fixed value for extinction. In three clusters, the LyC
flux calculated in this way would be approximately half of the original, while in two clusters we would obtain approximately twice the flux. Only in four \ion{H}{ii} regions would the ionizing photon output of the cluster be within 20\% of the previous one.

Another source of uncertainty in determining the escape fraction comes from the accuracy of the predicted ionizing photon fluxes from various theoretical stellar atmosphere models. This has been demonstrated by \citet{Oey97} and \citet{Voges08}, who determined the escape fraction from \ion{H}{ii} regions within the LMC using the same set of data, but various sets of atmosphere models. While \citet{Oey97} concluded that most of the regions are density bound, \citet{Voges08} determined the regions to be radiation bound using newer sets of atmosphere models. They also compared the ionizing photon output as a function of stellar spectral types for different sets of models in their Fig.~1. We used the BT-Settl2010 grid from \citet{Allard11} to estimate the $Q^0$ value of the stars to obtain consistent results because \textsc{ta-da} also uses these models to fit the stellar parameters. Additionally, we tested how the outcome of our estimates of the ionizing photon fluxes depends on the used stellar models. For this, we used the hot-star atmosphere model grids of \citet{Smith02}, which are based on models calculated with the WM-Basic code \citep{Pauldrach01}. The model atmosphere grids are calculated for the luminosity classes V and I for 5 different metallicities. Giant star models were not calculated explicitly and are just interpolations between the dwarf and supergiant models. 
We found that in all but two cases, the integrated LyC flux is higher when using the \citet{Smith02} models, by a medium factor of two. Only in the \ion{H}{ii} regions 17 and 19, which in our fitting are dominated by very hot stars ($T_{eff} > 50,000$~K),
is the ionizing photon flux about the same (within 5\%). 

In conclusion we can state that the examined approach is dominated by uncertainties in determining the spectral types of the stars within the \ion{H}{ii} regions and the photometric classification is not accurate enough. There are too many unknown or not precisely known parameters, such as the age of the stars, the extinction toward the stars, and the metallicity, to obtain a reasonable fit for the temperature and the surface gravity of the stars.
These are crucial to accurately determine their $Q^0$ values with broadband data alone, however. 
In a following project, the method introduced by \citet{Pellegrini12} might be employed to determine the leakage of ionizing photons from the objects studied here. Already existing narrowband data of NGC~300 in [\ion{O}{iii}] and [\ion{S}{ii}] emission lines taken with the Wide Field Imager (WFI) instrument at the MPG/ESO telescope can be used as a starting point. Alternatively, stellar spectra can be used for a more accurate stellar classification, as has already been done for the Magellanic Clouds (e.g., \citealt{Oey97,Voges08, Doran13}). It should be possible to obtain a reasonable estimate of the escape fraction of regions with low stellar densities by spectroscopically determining the temperatures of the hottest four to five stars within them because these dominate the total ionizing photon output.

\begin{acknowledgements} 
We thank the anonymous referee for useful comments and suggestions that helped improve the manuscript
This research was supported by the DFG cluster of excellence "Origin and Structure of the Universe".
NB is partially funded by a Royal Society University Research Fellowship.

\end{acknowledgements}

\end{document}